\documentclass[aps,pra,superscriptaddress,twocolumn,longbibliography,floatfix]{revtex4-1}

\usepackage{braket}
\usepackage{float}
\usepackage{graphicx}
\usepackage{amsmath}

\usepackage{units}
\usepackage{amsthm}
\usepackage{amssymb}
\usepackage{graphicx}
\usepackage{color}
\usepackage{bbold}
\usepackage[colorlinks=true, urlcolor=blue, citecolor=blue,linkcolor=blue,citebordercolor={1 0 0},linkbordercolor={0 0 1}]{hyperref}

\graphicspath{{./images/},{./imagesAppendix/}}

\theoremstyle{plain}

\theoremstyle{plain}

\ifx\proof\undefined
\newenvironment{proof}[1][\protect\proofname]{\par
	\normalfont\topsep6\p@\@plus6\p@\relax
	\trivlist
	\itemindent\parindent
	\item[\hskip\labelsep\scshape #1]\ignorespaces
}{%
	\endtrivlist\@endpefalse
}
\providecommand{\proofname}{Proof}
\fi
\theoremstyle{plain}

\theoremstyle{remark}

\makeatother

\providecommand{\factname}{Fact}
\providecommand{\theoremname}{Theorem}
\providecommand{\claimname}{Claim}
\providecommand{\lemmaname}{Lemma}
\providecommand{\definitionname}{Definition}

\theoremstyle{definition}

\usepackage{siunitx}
\usepackage{amssymb}
\usepackage{color}
\usepackage{tikz}
\usetikzlibrary{quantikz}
\usepackage{booktabs}

\definecolor{KB}{rgb}{0.4,0.3,0.9}

\definecolor{THc}{rgb}{0.9,0.3,0.2}

\definecolor{Change}{rgb}{0,0,0}

\begin{document}
% Define document title and author
\title{\textcolor{Change}{Noisy intermediate scale quantum simulation of time dependent Hamiltonians}}
\author{Jonathan Wei Zhong Lau}	
\email{e0032323@u.nus.edu}
\affiliation{Centre for Quantum Technologies, National University of Singapore 117543, Singapore}
\author{Kishor Bharti}
\email{kishor.bharti1@gmail.com}
\affiliation{Centre for Quantum Technologies, National University of Singapore 117543, Singapore}
\author{Tobias Haug}
\affiliation{QOLS, Blackett Laboratory, Imperial College London SW7 2AZ, UK}
\author{Leong Chuan Kwek}
\affiliation{Centre for Quantum Technologies, National University of Singapore 117543, Singapore}
\affiliation{National Institute of Education, Nanyang Technological University, 1 Nanyang Walk, Singapore 637616}
\affiliation{MajuLab, CNRS-UNS-NUS-NTU International Joint Research Unit, UMI 3654, Singapore}

% Write abstract here
\begin{abstract}

Quantum computers are expected to help us to achieve accurate simulation of the dynamics of many-body quantum systems. 
However, the limitations of current NISQ devices prevents us from realising this goal today. 
Recently an algorithm for performing quantum simulations called quantum assisted simulator has been proposed that promises realization on current experimental devices. 
In this work, we extend the quantum assisted simulator to simulate the dynamics of a class of time-dependent Hamiltonians. 
We show that the quantum assisted simulator is easier to implement as well as can realize multi-qubit interactions and challenging driving protocols that are difficult with other existing methods. 
We demonstrate this for a time-dependent Hamiltonian on the IBM Quantum Experience cloud quantum computer by showing superior performance of the quantum assisted simulator compared to Trotterization and variational quantum simulation. \textcolor{Change}{Further, we demonstrate the capability to simulate the dynamics of Hamiltonians consisting of 10000 qubits.}  Our results indicate that quantum assisted simulator is a promising algorithm for current term quantum hardware. 
\end{abstract}

\maketitle

%\noindent {\em Introduction.---} 
With the development of noisy intermediate-scale quantum (NISQ) devices, there is much interest in developing usable algorithms that can demonstrate ``quantum supremacy" for physically interesting problems on these devices. In particular, physicists hope to use these NISQ devices to probe the physics of many entangled particles, which are computationally expensive to simulate on classical computers~\cite{preskill2018quantum,deutsch2020harnessing}. The search for problems that these NISQ devices can realistically tackle are far-ranging, with notable fields being in quantum chemistry, solid-state physics and drug design. However, there still exist much pessimism regarding the potential of NISQ devices in the short term. For example, one of the most promising near-term algorithms in quantum chemistry is the variational quantum eigensolver (VQE) \cite{peruzzo2014variational,google2020hartree}, but it is hard to analyze the bounds of the performance of such algorithms~\cite{elfving2020will}, making it hard to determine if VQE will ultimately be feasible for practical use. One main problem is the barren plateau problem, which inhibits training via exponentially vanishing gradients in high dimensional variable space, that is preventing VQE from being more successful~\cite{mcclean2018barren,huang2019near,cerezo2020cost,wang2020noise}. Furthermore, these algorithms also lack a systematic approach to construct a sufficiently expressible and trainable Ansatz, with most current implementations resorting to hardware-efficient Ansätze~\cite{kandala2017hardware}. More physically inspired Ansatzes have been proposed, like Unitary Coupled Cluster Ansatz~\cite{taube2006new,lee2018generalized}, but these can be very complicated to implement.
    
As a systematic alternative to VQE, quantum assisted eigensolver (QAE)~\cite{bharti2020quantum} and iterative quantum assisted eigensolver (IQAE)~\cite{bharti2020iterative} have been recently proposed in literature. The classical optimization program corresponding to these algorithms is quadratically constrained quadratic program (QCQP) and thus well characterized, unlike VQE where the classical optimization step has not been well characterized. In particular, IQAE  provides a systematic method to build the Ansatz, bypasses the barren plateau problem, and also can be efficiently implemented on current hardware.
    
The methods described in QAE and IQAE can be easily applied to the problem of simulating quantum dynamics as well. The task of simulating quantum dynamics is extremely challenging for a classical computer, and Feynman proposed that this would be one of the areas where quantum computers would show clear advantages over classical computers~\cite{feynman1982simulating}. 
    
Using the same ideas of Ansatz generation in IQAE, quantum assisted simulator (QAS)~\cite{bharti2020quantum2,haug2020generalized} has been proposed as a NISQ compatible algorithm to simulate quantum dynamics. 
Due to its efficient implementation, Ansatz construction technique, and lack of a quantum-classical training feedback loop, QAS promises to deal with the problems that are inherent in other variational algorithms that have been proposed in the literature for simulating quantum dynamics on NISQ devices. 
    
In this work, we build on the work of QAS and its successor, generalized quantum assisted simulation (GQAS) \cite{haug2020generalized} and apply it to a certain class of time-dependent Hamiltonians. We also compare it to two alternatives, variational quantum simulator (VQS) and Trotterization, on a real quantum processor and show that QAS has superior performance on the NISQ device compared to the other two methods.

\medskip
{\noindent {\em Background of QAS---}}
% \section*{Background of QAS} \label{background}
	In quantum simulation we aim to simulate the time-independent Schr\"{o}dinger equation with Hamiltonian $H$ for a closed system, represented by $\ket{\phi(t)}$:
	\begin{gather}
	    \frac{\text{d}\ket{\phi(t)}}{\text{d}t}=-i H \ket{\phi(t)}.
	\end{gather}
	The QAS algorithm proposes a hybrid quantum-classical Ansatz
	\begin{equation}
	    \ket{\phi(\Vec{\alpha}(t))} = \sum_i \alpha_i(t) \ket{\chi_i}.
	\end{equation}
    The coefficients $\alpha_i(t)$ are variational parameters and stored on a classical computer. The $\ket{\chi_i}$ are fixed quantum states that can be produced on a quantum computer and called the K-moment states. These states are generated from the Hamiltonian and are defined in Appendix \ref{Appendix_C}. 
    
    Using either the Dirac and Frenkel variational principle \cite{dirac1930note,frenkel1934wave} or McLachlan's variational principle \cite{mclachlan1964variational}, we obtain the evolution equation
    \begin{gather}
        \sum_j \mathcal{E}_{i,j}\Dot{\alpha}_j = -i \mathcal{C}_i, \label{evolution_eqn}
    \end{gather}
    where
    \begin{gather}
        \mathcal{C}_i = \sum_j \mathcal{D}_{i,j}\alpha_j (t),\\
        \mathcal{E}_{i,j} = \braket{\chi_i|\chi_j},\label{eq:overlap_E}\\
        \mathcal{D}_{i,j} \equiv \bra{\chi_i}H\ket{\chi_j}.\label{eq:overlap_D}
    \end{gather}
    Thus the only values that we require to be calculated on the quantum computer are \eqref{eq:overlap_E} and \eqref{eq:overlap_D}.
    Most notably, we stress that these overlaps have to be calculated only once, after which the quantum part of the algorithm is finished. There is no quantum-classical feedback loop necessary to optimise parameters.
    Furthermore, by using methods found in \cite{mitarai2019methodology}, usually the form of the Hamiltonian allows us to measure these overlaps without using Hadamard tests or ancilla qubits, which are problematic in NISQ devices. If the Hamiltonian is a linear combination of tensored Pauli operators, the elements can be directly inferred from the measurement
    in the corresponding computational basis. Then we use a classical computer to update the $\alpha_i$ parameters for every step forward in time according to the equations:
    \begin{gather}
        \sum_j \Dot{\alpha_j}\braket{\chi_i|\chi_j}=-i\sum_j\alpha_j \braket{\chi_i |H| \chi_j}\label{evolutioneqn}\\
        \alpha_j(t + \delta t)=\alpha_j(t) + \Dot{\alpha}_j \delta t\label{evolutioneqn2}.
    \end{gather}
    This implies that the step size $\delta t$ of the evolution in this algorithm can be made arbitrarily small with no additional cost on the quantum computer. Once we calculate the classically hard to obtain overlap values on the quantum computer, we can freely reduce the step size in our evaluation on the classical computer until we obtain a desired accuracy.
    This makes the algorithm very advantageous in situations where we expect the quantum dynamics to be changing very fast in time, as in those situation reducing the step size will be very important in capturing the accurate dynamics. 
    
    \medskip
{\noindent {\em \textcolor{Change}{Comparison with existing approaches---}}}
    The QAS approach to simulate quantum dynamics has intuitive advantages over the current alternatives. 
    \textcolor{Change}{The Trotterization method to simulate quantum dynamics utilizes the Trotter-Suzuki decomposition of the unitary time evolution operator into small discrete steps. Each step is individually made up of efficiently implementable quantum gates, which can be run on the quantum computer~\cite{lloyd1996universal,lanyon2011universal,peng2005quantum,barends2016digitized,barends2015digital,martinez2016real,sieberer2019digital}.} However, this is currently thought to be unfeasible for NISQ devices for long time evolution, accounting for the number of gates that they would require, and we would  probably need to wait for fault tolerant quantum computers before Trotterization is able to be implemented with high fidelity~\cite{poulin2014trotter}. 
    Alternate variational quantum algorithms have been proposed in the literature, such as variational quantum simulation (VQS)~\cite{endo2020variational,yuan2019theory,li2017efficient} and subspace variational quantum simulator (SVQS)\cite{heya2019subspace}, and attempts have been made to apply VQS to some simple chemical dynamics \cite{lee2021simulating}. Both algorithms require the use of a quantum-classical feedback loop. The VQS algorithm dynamically updates the parameters of a parametric quantum circuit with a quantum-classical feedback loop, such that the evolution of a parameterized quantum state mimics the real quantum evolution, while SVQS uses this quantum-classical feedback loop to find a certain number of excited states of a given Hamiltonian and to establish a mapping between the computational basis and the energy eigenbasis of the Hamiltonian. However both algorithms share similar problems to VQE like the barren plateau problem, and both algorithms do not have a systematic way to generate a parameterized Ansatz, although adaptive Ansatz generating procedures have been recently proposed \cite{yao2020adaptive}. VQS suffers from requiring Hadamard tests as part of its measurement scheme, which can be expensive to implement on current devices. It is possible to replace the Hadamard tests required in VQS with direct measurements, but these methods will have scaling problems, as the number of circuits that needs to be evaluated scales exponentially with the number of controlled gates in the Ansatz \cite{mitarai2019methodology}. Methods like variational fast forwarding (VFF)\cite{cirstoiu2020variational,commeau2020variational} have also been proposed where they seek to approximately diagonalize short-time dynamics to enable longer-time simulations. VFF has also been tested for small systems on current hardware and appears to be quite successful at simulating small systems so far. However, the diagonalization step of VFF is still not studied well, and the no fast-forwarding theorem suggests that not all Hamiltonians will be able to be accurately diagonalized with a reasonable amount of gates and circuit length, and the optimization step of the cost function in VFF might be too difficult to carry out efficiently. Hybrid approaches, like combining Trotterization and variational methods, have also been proposed \cite{benedetti2020hardware}, but they also suffer from the problems associated with VQS dealing with optimizing over a quantum-classical feedback loop, like the barren plateau problem.

\begin{table*}[]
\centering
\caption{Comparison of different methods for simulating quantum dynamics on NISQ devices.}
\resizebox{\textwidth}{!}{%
\begin{tabular}{|c|c|c|c|c|}
\hline
Criteria & Quantum Assisted Simulation & Variational Quantum Simulation & Variational Fast Forwarding & Trotterization \\ 
& (QAS)\cite{bharti2020quantum2,haug2020generalized} & (VQS) \cite{endo2020variational,yuan2019theory,li2017efficient}& (VFF)\cite{cirstoiu2020variational,commeau2020variational} & \\
\hline
\begin{tabular}[c]{@{}c@{}}Circuit \\ Complexity\end{tabular} & \begin{tabular}[c]{@{}c@{}}Usually only requires \\ measurements of expectations \\ of Pauli strings, which \\ avoids Hadamard tests. \\ Avoids complicated \\ controlled-unitary gates.\end{tabular} & \begin{tabular}[c]{@{}c@{}}Usually requires Hadamard \\ tests and complicated \\ controlled-unitary gates.\\ Procedures exist to avoid \\ Hadamard tests, but have \\ scaling problems, as number\\  of circuits that needs \\ to be measured grows\\ exponentially with number \\ of controlled gates in \\ the Ansatz.\end{tabular} & \begin{tabular}[c]{@{}c@{}}Usually requires relatively \\ complicated circuits.\\ No Fast Forwarding\\ theorem suggest that not all\\ Hamiltonians can be \\ accurately diagonalized with\\ a reasonable amount of \\ gates and circuit length.\end{tabular} & \begin{tabular}[c]{@{}c@{}}Circuit length grows \\ linearly with number \\ of steps.\end{tabular} \\ \hline
\begin{tabular}[c]{@{}c@{}}Usage of \\ Quantum\\ Computer\end{tabular} & \begin{tabular}[c]{@{}c@{}}Minimizes use, no classical-\\ quantum feedback loop. All\\ circuits can be measured at \\ the start, and further evolution \\ is solely done on classical\\ computer.\end{tabular} & \begin{tabular}[c]{@{}c@{}}Requires classical-quantum \\ feedback loop. Constantly \\ needs use of quantum \\ computer for every timestep \\ of evolution.\end{tabular} & \begin{tabular}[c]{@{}c@{}}Requires classical-quantum \\ feedback loop at the start. \\ However, once first step is \\ done, only requires minimal \\ usage.\end{tabular} & \begin{tabular}[c]{@{}c@{}}Requires use for every \\ step, but minimal usage\\ as only one circuit is \\ required at every step.\end{tabular} \\ \hline
\begin{tabular}[c]{@{}c@{}}Barren\\ Plateau\\ Problem\end{tabular} & None & \begin{tabular}[c]{@{}c@{}}May exist.\end{tabular} & \begin{tabular}[c]{@{}c@{}}May exist.\end{tabular} & None \\ \hline
\begin{tabular}[c]{@{}c@{}}Multi\\ Qubit\\ Interactions\end{tabular} & \begin{tabular}[c]{@{}c@{}}Easy to simulate, reduces\\ to a Pauli string measurement\end{tabular} & \begin{tabular}[c]{@{}c@{}}Hard to simulate, requires\\ mult-qubit gates.\end{tabular} & \begin{tabular}[c]{@{}c@{}}Possible to simulate,\\ part of diagonalization \\ process.\end{tabular} & \begin{tabular}[c]{@{}c@{}}Hard to simulate,\\ requires multi-\ qubit gates.\end{tabular} \\ \hline
\begin{tabular}[c]{@{}c@{}}Ansatz \\ Generation\end{tabular} & Systematic. & No known systematic method. & No known systematic method. & No need for Ansatz. \\ \hline
\begin{tabular}[c]{@{}c@{}}Discretization\\ cost\end{tabular} & \begin{tabular}[c]{@{}c@{}}Can be freely adjusted\\ with no additional cost\\ on quantum computer.\end{tabular} & \begin{tabular}[c]{@{}c@{}}Cost on quantum computer\\ increases linearly with \\ number of steps.\end{tabular} & \begin{tabular}[c]{@{}c@{}}No discretization\\ (Approximate \\ diagonalization of \\ Hamiltonian)\end{tabular} & \begin{tabular}[c]{@{}c@{}}Cost on quantum\\ computer increases\\ linearly with number\\ of steps.\end{tabular} \\ \hline
\end{tabular}%
}\label{tablecomparison}
\end{table*}

    The advantage of QAS as opposed to VQS is that not only does it circumvent the barren plateau problem by construction, QAS does not require complicated measurement protocols involving Hadamard tests and controlled unitaries, which are difficult to implement and introduce on current NISQ devices. Compared to SVQS, QAS requires much shorter circuit lengths and less gate operations, which give less room for noise. Further, it is not limited to finding the dynamics in a low-energy subspace and it does not rely on the problems associated with VQE that SVQS relies on to find the excited states.  
    As further advantage, the timesteps can be made arbitrarily small without any additional computional cost on the quantum computer, in contrast to Trotter-based methods or VQS.
    QAS also promises to minimize the usage of the quantum computer and reduce circuit complexity as compared to the other methods. A comparison of various methods for quantum simulation is found in Table \ref{tablecomparison}.

% % Main Part
% \section{Approach}\label{approach}
\medskip
{\noindent {\em \textcolor{Change}{Approach for time-dependent dynamics---}}}
	Now we want to simulate time-dependent Hamiltonian of the form $\sum_i f_i(t)H_i$. For arguments sake, let us consider a Hamiltonian of the form:
	\begin{gather}
	    H(t) = H_s + f(t)H_t,\\
	    H_s = \sum_i \beta_i U_i,\label{ui}\\
	    H_t = \sum_j \gamma_j V_j,\label{uj}
	\end{gather}
    where $f(t)$ is some `classical' function (eg $\sin(t)$), and where each unitary $U_i$, $V_j$ in $H_s$ and $H_t$ acts non-trivially on at most $\mathcal{O}(poly(logN))$. If the unitaries are tensored Pauli matrices, we do not need the $\mathcal{O}(poly(logN))$ constraint. In the theory of quantum control~\cite{d2007introduction,dong2010quantum,dirr2008lie}, the possible evolution operators that can be achieved as a solution of the Schr\"{o}dinger equation are of the form:
    \begin{gather}
        \exp\left(-i H_\text{eff} t \right).
    \end{gather}
    $H_{\text{eff}}$ has a specific structure, where it can be written as a linear combination of all the possible nested commutators of $H_s$ and $H_t$, $\{ H_s, [H_s,H_t], [H_s,[H_s,H_t]], \dots \}$. The set of operators that can be formed from these nested commutators is known as the dynamical Lie algebra of the system, and it can be shown that the set of reachable states (states that can be obtained via evolution under the Hamiltonian) is connected with this dynamical Lie algebra \cite{dirr2008lie}. This essentially translates the time-dependent problem into a time-independent one as $H_{\text{eff}}$ is a effective time-independent Hamiltonian that evolves the system in the same way as the original time-dependent Hamiltonian.
    \begin{gather}
        H_{\text{eff}} = \sum_m c_m G^\prime_m, \\
        G^\prime_m \in \{ H_s, [H_s,H_t], [H_s,[H_s,H_t]], \dots \}.
    \end{gather}
    The coefficients $c_m$ are dependent on the exact form of $f(t)$ and requires us to solve the time-dependent Schr\"{o}dinger equation exactly or numerically. This limits the number and complexity of systems that we are able to write down effective Hamiltonians for. However, in the QAS framework, we are interested in using the structure of the Hamiltonian to build our Ansatz. Since we know the terms that are in $H_{\text{eff}}$, we can use this to help construct our Ansatz.
    However, these $G^\prime_m$ are not guaranteed to be unitary. For example, if we consider a 3 qubit system with the Hamiltonians: 
    \begin{gather}
        H_s = (Z_1 Z_2 + Z_2 Z_3),\\
        H_t = X_2.
    \end{gather}
    The set of possible operators in $H_{\text{eff}}$ is then
    \begin{gather}
        \{X_2, (Z_1 Z_2 + Z_2 Z_3), (Z_1 Y_2 + Y_2 Z_3), Z_1 X_2 Z_3\}.
    \end{gather}
    As can be seen, the 3rd and 4th terms in this set are not unitary. However, all the terms in this set can be written as linear combination of unitaries, which we show in the example below. This allows us to form another set by decomposing all those non-unitary terms in the original set into unitary terms, and generating a new set with the terms from the original set and these unitary terms. We add in the identity operator, and call this the extended set of permitted operators in $H_{\text{eff}}$

    \begin{gather}
        S_{ext} = \{I\}\bigcup \{O_i\} \bigcup \{[O_i,O_j]\}\bigcup \{[O_i,[O_j,O_k]]\}\bigcup \dots\\
            O_i,O_j,O_k,\dots \in \{U_i\}\bigcup\{V_j\},
    \end{gather}
    
    where $U_i$ and $V_j$ are the unitary terms in the Hamiltonian, defined in equations \ref{ui} and \ref{uj}. In our example, this extended set looks like: 
    
    \begin{gather}
        S_{ext} = \{I,X_2, Z_1 Z_2, Z_2 Z_3, Z_1 Y_2 , Y_2 Z_3, Z_1 X_2 Z_3\}.
    \end{gather}
    
    We then use the terms in this set as a basis for the K-moment expansions. Given a set of unitaries $U^\prime_{i}\in\{S_{ext}\}\label{1_moment_new_2},$ a positive integer K and some initial quantum state $\ket{\psi}$, the K-moment states is the set of quantum states of the form $\mathbb{S}_K=\{ U^\prime_{iK}\dots U^\prime_{i2}U^\prime_{i1} \ket{\psi} \}$, denoted by $\mathbb{S}_K$. The cumulative K-moment states $\mathbb{CS}_K$ is defined to be $\mathbb{CS}_K \equiv \bigcup _{j=0}^K \mathbb{S}_j$. We define our hybrid Ansatz as:
    
    \begin{gather}
	    \ket{\phi(\Vec{\alpha})}_K = \sum_{\ket{\chi_i} \in \mathbb{CS}_K} \alpha_i \ket{\chi_i}\label{ansatz_new},
	\end{gather}
    for any arbitrary quantum state $\ket{\psi}$.  
    
    Now we look at Eqs. \eqref{evolutioneqn} and \eqref{evolutioneqn2} again, but now with the time-dependent Hamiltonian. Equation \ref{evolutioneqn} is rewritten as
    \begin{gather}
        \sum_j \Dot{\alpha_j}\mathcal{E}_{i,j}=-i\sum_j\alpha_j \mathcal{D}_{i,j}\label{evolutioneqnnew},
    \end{gather}
    where $\mathcal{E}_{i,j}=\braket{\chi_i|\chi_j}$, and $\mathcal{D}_{i,j}=\braket{\chi_i |H(t)| \chi_j}$. Now, we note that to calculate $\braket{\chi_i |H(t)| \chi_j}$, we do not need to continuously use the quantum computer for every different $t$. To see this, we rewrite $\braket{\chi_i |H(t)| \chi_j}$:
    \begin{gather}
        \braket{\chi_i |H(t)| \chi_j} = \braket{\chi_i |H_s| \chi_j} + f(t)\braket{\chi_i |H_t| \chi_j}.
    \end{gather}
    As can be seen, since the time-dependent function $f(t)$ can be brought outside of the braket, and $H_t$ does not change, we only need to use the quantum computer once to calculate the $\braket{\chi_i |H_s| \chi_j}$ and $\braket{\chi_i |H_t| \chi_j}$ terms. With those overlaps, a classical computer can easily calculate $\braket{\chi_i |H(t)| \chi_j}$ when given $f(t)$. Circuits that perform the measurements of these quantities that do not need a Hadamard test are given in section \ref{Appendix_B}.
    
    This showcases the main advantage of QAS as it is highly efficient in using the quantum computer. It has to be only run at one point to calculate overlaps which can be performed in parallel, without the need of feedback loops.

    Furthermore, one thing to note is that although in this paper we only considered a single static Hamiltonian and time-dependent Hamiltonian term, this method is easily extended to multiple time-dependent Hamiltonian terms that all have different dependencies on time,  $H(t) = \sum_i f_i(t)H_i$. The algorithm is as follows
    \begin{enumerate}
        \item From the original Hamiltonian $H(t)$ construct $S_{ext}$.
        \item Choose an efficiently implementable initial state $\ket{\psi}$.
        \item Form the K-moment states $\ket{\chi_i}$ for some K$>$0, then construct the Ansatz with arbitrary $\alpha_i$ values.
        \item Use the quantum computer to calculate the overlap values $\mathcal{E}_{i,j}$ and $\mathcal{D}_{i,j}$ once. 
        \item Evolve the state forward in time using a classical computer.
    \end{enumerate}

    In a similar manner, this procedure can be extended to deal with time-dependent Hamiltonians in the open system case, described in Appendix \ref{opensystemextension}. 
    
    \textcolor{Change}{One possible hurdle to overcome is that our method might have scaling problems. This issue and a more detailed scaling analysis is further elaborated in Appendix \ref{appendix:expressibility} and \ref{appendix:scaling}. }

    \begin{figure*}
        \centering
        \includegraphics[width=6in]{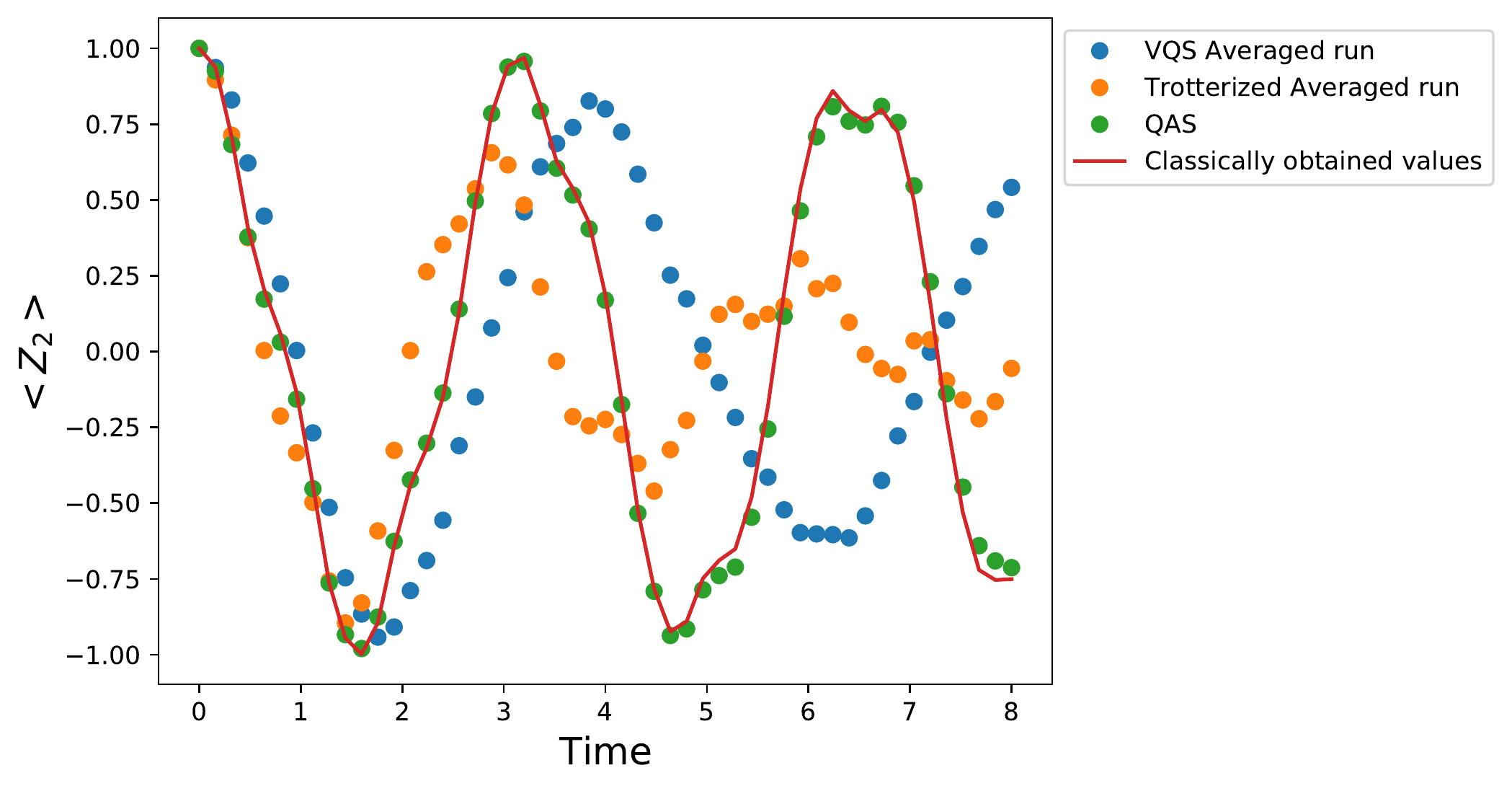}
        \caption{Time evolution of Hamiltonian Eq.\ref{eq:Hamiltonian} simulated on IBM quantum processor \emph{ibmq\_rome} for QAS, VQS and Trotterization. We evolved dynamics of the expectation value of $\braket{Z_2}$ in 50 steps between $t=0$ to $t=8$ for all the methods. The data for VQS and Trotterization was averaged over 3 runs. \textcolor{Change}{The error mitigation scheme for this plot is described in Appendix \ref{appendix:errormitigation}.} The results for $\braket{Z_2}$ with QAS are in good agreement with exact results, whereas VQS and Trotterization notably deviate. The initial state was created with the circuit shown in Appendix \ref{randomcircuit} using randomized parameters.}
        \label{fig:realqc}
    \end{figure*}

    \medskip
{\noindent {\em Examples---}}
    We simulate the QAS algorithm on some examples of 1 qubit, 3 qubit, 7 qubit and 11 qubit closed systems under different time-dependent Hamiltonians, and an open system of 6 qubits. The results and details are shown in Appendix \ref{QAS_examples}. From our simulations, we show that QAS is indeed capable of capturing the expected dynamics of time-dependent Hamiltonians. More importantly, we show that in this framework, we can easily simulate multiple-qubit interactions in the Hamiltonian, and since we can freely adjust the timestep on the classical computer, we can  capture the dynamics of high-frequency time-dependent Hamiltonians without incurring a higher cost on the quantum computer.
    
    We also benchmark the QAS algorithm with an experimental demonstration on a quantum superconducting processor, and compare it to VQS and Trotterization. We utilize the IBM cloud quantum computer provided by IBM quantum experience.
    
    % All our data was obtained on December 1-3 2020, on 
    We ran the algorithms on the 5-qubit quantum computer \emph{ibmq\_rome}, available through IBM Quantum Experience. Here, we considered a 2 Qubit example:
    \begin{gather}\label{eq:Hamiltonian}
        H(t) = X_1 X_2 + \sin (2 \pi t) Y_2.
    \end{gather}
    The extended set of permitted operators that we will use for QAS is given as
    \begin{gather}
        S_{ext} = \{I,X_1 X_2, Y_2, X_1 Z_2\}.
    \end{gather}
    This gives us our Ansatz for QAS (using $K=1$ for the $K$-moment expansion):
    \begin{gather}
        \ket{\phi(\Vec{\alpha})}_{\text{QAS}} = \alpha_1 \ket{\phi_0} + \alpha_2 X_1X_2\ket{\phi_0}\notag\\
        + \alpha_3 Y_2\ket{\phi_0} + \alpha_4 X_1Z_2\ket{\phi_0}.
    \end{gather}

    In this example, the initial state $\ket{\phi_0}$ that we used for QAS, VQS and Trotterization is a 2-qubit circuit with randomized parameters, generated with a layer of single qubit rotations, and an entangling gate between the two qubits (see Appendix \ref{randomcircuit}). QAS requires $\approx 10$ circuits to be evaluated on the quantum computer (1 job = running 1 circuit). By comparison, Trotterization requires $50$ circuits, and VQS requires $\approx 500$ circuits to be evaluated on the quantum computer. For VQS, we used an Ansatz with similar terms
    \begin{gather}
        \ket{\phi(\theta_1,\theta_2,\theta_3)}_{\text{VQS}} = e^{-i \hbar \theta_1 X_1 X_2}e^{-i \hbar \theta_2 Y_2}e^{-i \hbar \theta_3 X_1 Z_2}\ket{\phi_0}.\label{vqsansatz}
    \end{gather}
The Trotterization of the time evolution operator is decomposed as
\begin{gather}
         e^{-i \hbar \tau H(t)} \approx \prod_{i=1}^N e^{-i \hbar \delta t_i X_1 X_2}e^{-i \hbar \delta t_i \sin (2 \pi t_i) Y_2}\,,
\end{gather}
with  $\sum_{i=1}^N \delta t_i = \tau$ and $\sum_{j=1}^i \delta t_j = t_i$.
For all the algorithms, we ran it for 50 steps, from $t=0$ to $t=8$. The results are shown in Figure \ref{fig:realqc}.
     As can be seen, QAS is successful in accurately simulating the dynamics, with results very close to the expected results, while being much more efficient in the use of the quantum computer as compared to alternatives. We ran the VQS and Trotter algorithm on the quantum computer 3 times and obtained an averaged value for the data for VQS. For all the runs of the VQS and Trotter algorithm, the results do not reproduce the correct results as well as QAS. The depth of the circuits required for Trotterization rapidly grows too deep to be usable. 
     
     As we would like to emphasize, this was done with a much more efficient use of the quantum computer. As compared to VQS, which utilizes a sort of quantum-classical feedback loop, QAS does not require any quantum-classical feedback loop. We only required around 30 minutes of usage of the quantum computer for our run of QAS ($\approx 10$ circuits). Trotterization required around 2 hours  ($\approx 50$ circuits). Compared to them, VQS required almost 3 days for us to complete 1 run ($\approx 500$ circuits). Even with dedicated access to a quantum computer, this is slightly unreasonable, especially as the time will scale linearly with discretization of the time.
     
    \begin{figure*}
        \centering
        \begin{tikzcd}
            \lstick{$\ket{0}$}&\gate{H}&\gate{X}&\ctrl{1}&\ctrl{2}&\qw&\ctrl{2}&\gate{X}&\gate{H}&\meter{}\\
            \lstick{$\ket{\phi_0}$}&\qw&\gate[2]{R_{xz}(\theta_3)}&\gate{X}&\qw&\qw&\qw&\qw&\qw&\qw\\
            \lstick{$\ket{\phi_0}$}&\qw& &\qw&\gate{Z}&\gate{R_y(\theta_2)}&\gate{Y}&\qw&\qw&\qw
        \end{tikzcd}
        \caption{Example circuit for measuring overlaps needed for VQS for the evolution of a quantum state $\ket{\psi}$ in time. Requires controlled unitaries for the Hadamard test. The $\theta$s are the variational parameters in the Ansatz for VQS, specified in equation \ref{vqsansatz}}
        \label{fig:VQS_circuit}
    \end{figure*}
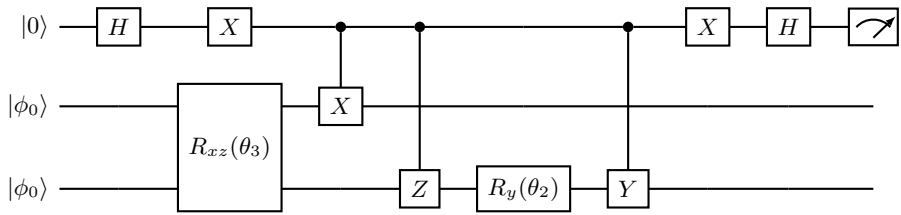
    
    \begin{figure*}
        \centering
        \begin{tikzcd}
            \lstick{$\ket{\phi_0}$}&\gate{H}&\qw&\meter{}\\
            \lstick{$\ket{\phi_0}$}&\gate{S^\dagger}&\gate{H}&\meter{}
        \end{tikzcd}
        \caption{Example circuit for QAS needed for measuring overlaps for simulating the evolution of an initial quantum state $\ket{\psi}$ in time. Only single-qubit rotations and measurements in the computational basis are required.}
        \label{fig:QAS_circuit}
    \end{figure*}
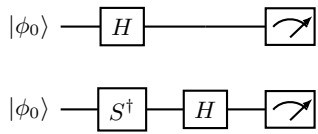

     In QAS, the circuits that need to be implemented are also much simpler than those in VQS. An example of a typical circuit that has to be calculated for this case is given in Fig.\ref{fig:VQS_circuit}. As can be seen, this circuit is relatively long and complicated, requiring controlled unitary gates and parameterized rotation gates. Furthermore, due to the linear architecture that is common for the superconducting qubit quantum computers (like those on the IBM cloud that we accessed), the presence of these 2-qubit rotation gates and controlled unitaries will require SWAP gates to be implemented in general. All these sort of gate operations are very costly and prone to errors. Compare this to a typical qubit needed for QAS given in Fig.\ref{fig:QAS_circuit}, which is short and does not require many complicated 2-qubit gates or SWAP operations to be implemented. 
     
     Also, there seems to be a systematic error inherent in VQS when subjected to the errors of a real quantum computer, which causes the time evolution of the expectation value to lag behind the correct result and increases the period of the dynamics. To illuminate the origin of this effect, we constructed and simulated a simple noise model on a classical computer, which comprised of depolarizing error with a noise parameter $\lambda$:
     \begin{gather}
    E_{\text{depolarizing}}(\rho) = (1-\lambda)\rho + \lambda \text{Tr}[\rho^2]I.
\end{gather}
     The depolarizing error is the probability that a single qubit, after going through a gate, will be replaced by the completely mixed state. We tested VQS with different values for the noise parameter. The results are shown in Appendix \ref{appendix:noise}. As we increase the noise parameter,the lag of VQS increases, indicating that VQS in the presence of noise tends to cause the dynamics to move slower than they should. This leads to a systemtatic overestimation of the oscillation periods in the dynamics. If we look at the VQS equation found in  \cite{endo2020variational}:
     \begin{gather}
         \mathcal{M}\Dot{\Vec{\theta}} = \Vec{V},
     \end{gather}
     we note that both $\mathcal{M}$ and $\Vec{V}$ are affected by depolarizing noise, which tends to reduce the expectation values closer to 0. If $\mathcal{M}$ and $\Vec{V}$ are affected equivalently, the noise should cancel out. However in this 2-qubit example, the average circuit depth for an element in $\mathcal{M}$ is shorter than the average circuit depth for an element in $\Vec{V}$ by around $33\%$ (due to some gates being redundant when compiled). Thus, $\Vec{V}$ tends to experience more depolarizing error, which in general reduces the expectation value of $\Vec{V}$ closer to 0. Thus, the norm of $\Vec{V}$ is reduced more than of $\mathcal{M}$. This results in smaller time derivatives for $\Vec{\theta}$, which results in smaller frequencies, and systematic error causing slower dynamics. However in QAS, our $\mathcal{E}$ and $\mathcal{D}$ matrices are of the same depth, thus we do not see problems of this kind arising.

     We could have implemented the methods in \cite{mitarai2019methodology} to eliminate the need for the Hadamard test for VQS, but that would have quadrupled the amount of circuits we would have needed to have run on the quantum computer, as that method scales exponentially with the complexity of the Ansatz and size of the system. Considering the high amount of circuits VQS already needs, a quadrupling of the amount of circuits would have been unfeasible for us to run in a reasonable amount of time with our current resources and the lack of allocated dedicated time to run our circuits on the cloud IBM service.
    
\begin{figure}
    \centering
    \includegraphics[width=0.49\textwidth]{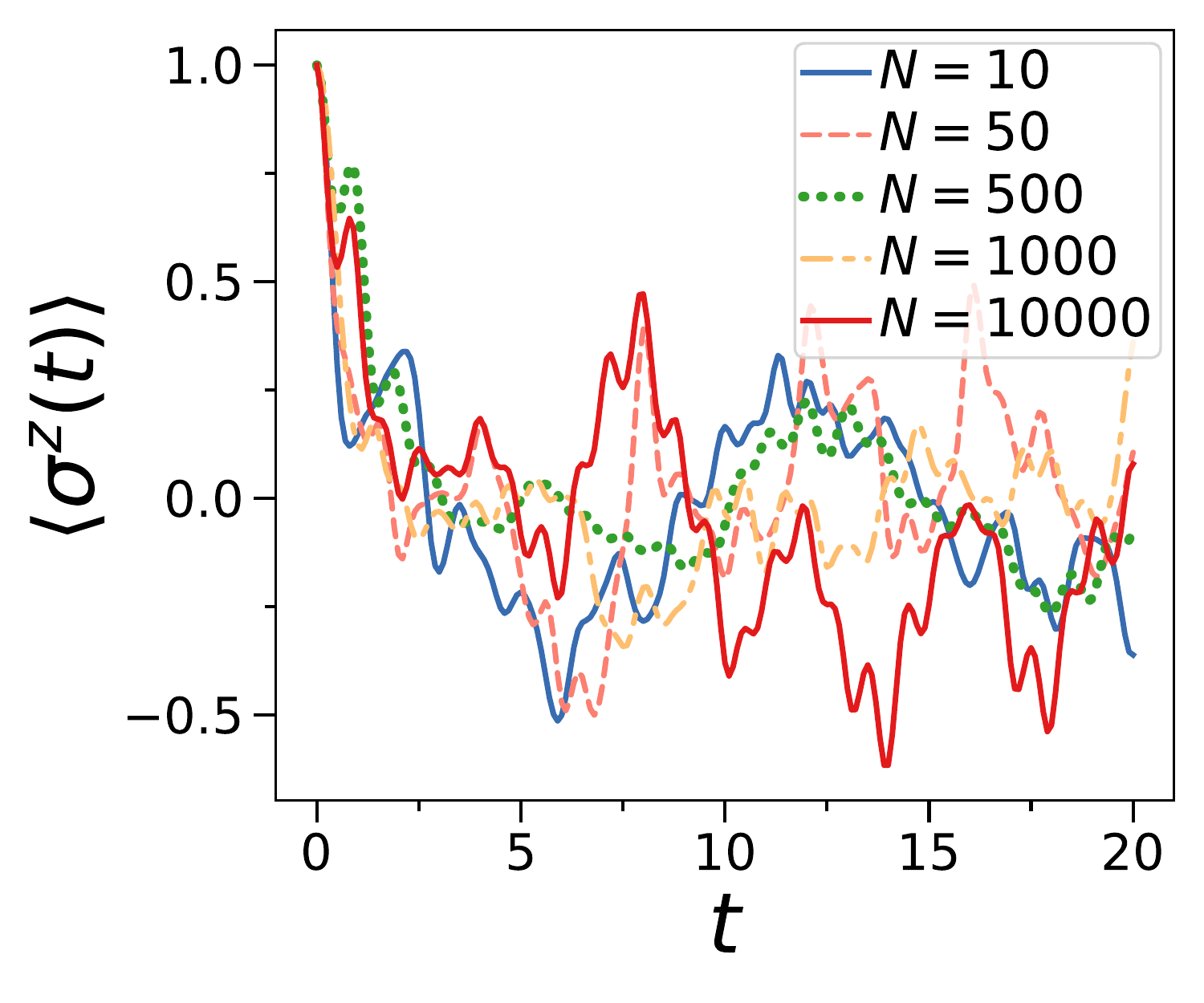}
    \caption{Time evolution of a Hamiltonian consisting of $r=8$ random $N$-qubit Pauli strings with time-dependent coefficients. We evolve the quantum state $\ket{0}^{\otimes{N}}$ for various number of qubits $N$. We use the maximal number of $K$-moment states for the evolution, consisting of $2^r=256$ quantum states, which gives the exact dynamics.}
    \label{fig:multiqubit}
\end{figure}

\textcolor{Change}{Our approach can tackle problems consisting of a large number of qubits $N$, which are difficult for both variational quantum simulation methods based on parameterized quantum circuits, as well as tensor network methods.
In Fig.\ref{fig:multiqubit}, we show the time evolution with a time-dependent Hamiltonian composed of $r$ different random Pauli strings $H=\sum_{i=1}^r \sin(\pi t+\phi_i) P_i$, where the Pauli strings $P_i=\otimes_{j=1}^N \boldsymbol{\sigma}_j$ consist of $N$ tensored Pauli operators $\boldsymbol{\sigma}_j\in\{I,\sigma^x,\sigma^y,\sigma^z\}$ and the phase $\phi_i$ is randomly chosen $\phi_i=\{0,\pi/2\}$.  
The initial state $\ket{\psi}=\ket{0}^{\otimes{N}}$ is the $N$-qubit product state with all zeros. To generate the hybrid Ansatz, we use all cumulative $K$-moment states, which take maximally $2^r$ different states. We show the simulated dynamics using our method for up to $N=10000$ qubits. 
While we can numerically demonstrate the dynamics by using a product state $\ket{0}^{\otimes{N}}$ as initial state, the problem becomes intractable for classical computers when the initial state $\ket{\psi}$ is a highly entangled quantum state.
Such highly entangled states can be generated on quantum computers~\cite{arute2019quantum} and their dynamics could be then simulated using our method. }

\medskip
{\noindent {\em Discussion and Conclusion---}}
In this work, we provide an extension of QAS for simulating the dynamical evolution of a quantum system under a time-dependent Hamiltonian. The advantages of this algorithm when compared to other quantum simulation methods are summarized in the following.

Firstly, our algorithm provides a systematic method to select the Ansatz, as opposed to current implementations of VQS. We use the structure of the Hamiltonian given to generate a set of operators, which are all unitary (the extended set of permitted operators). This allows us to build up a linear combination of quantum states, where the quantum states are the set of $K$-moment states generated using the operators in the extended set of permitted operators. The coefficients of these quantum states are complex numbers in general, and are the variational parameters that will be used to simulate the evolution of the state.

Secondly, in our framework, if we can express the Hamiltonian as a linear sum of Pauli strings, we do not require complicated circuits to be measured on the quantum computer. This is in contrast to VQS, which requires complicated Hadamard tests and controlled-unitary gates that can only be circumvented with methods that scale exponentially as the Ansatz complexity increases. It also stands in contrast to SVQS, which requires large unitaries that involve many 2-qubit gates over different qubits that is expensive to implement on NISQ devices with limited topology between qubits. 
Further, we find that VQS suffers from a systematic lag in the time evolution, which we can explain with depolarizing noise that affects the measurement circuits. QAS possesses short measurement circuits and does not suffer this problem.

Thirdly, QAS is often more efficient in its usage of the quantum computer as compared to VQS and Trotterization, as the number of circuits it needs to calculate for the entire procedure is usually much less. QAS does not require a classical-quantum feedback loop, and only utilizes the quantum computer at the start to calculate overlap values of certain quantum states. This step can be done in a parallel manner, which can further speed up calculations. Once these overlap values are calculated, the job of the quantum computer is done, and we solely rely on the classical computer to simulate the evolution of the state, no longer requiring evaluations on the quantum computer. This implies that the step size of the evolution can be made arbitrarily small with no additional cost on the quantum computer. Thus, it should have an advantage in simulating high-frequency time-dependent problems. This stands in contrast to VQS and Trotterization, where the computation cost on the quantum computer increases linearly with number of steps to be simulated.

Fourthly, we avoid the barren plateau problem that is present in variational algorithms that require finding gradients.

Lastly, in our framework it is easy to simulate multi-qubit interactions, which is hard for both QAS/SVQS and Trotterization as they would require implementation of controlled multiple-qubit rotation gates. 

However, there are still many problems with this approach. One cause for concern is that as the size of the system increases, and the complexity of the Hamiltonian increases, large $K$ values to generate $K$-moment states are needed to generate a large enough space of states to explore. \textcolor{Change}{This is also dependent on the number of states in $S_{ext}$, for the method described in this paper. The size of $S_{ext}$ depends on the Hamiltonian chosen, and we expect it to grow exponentially with the number of non-commuting terms in the Hamiltonian. This will increase the number of overlap terms the quantum computer has to calculate, and increase the size of the matrix equations the classical computer has to solve in equation \ref{evolutioneqnnew}. This fundamental challenge of finding an appropriate expressible Ansatz is encountered both in QAS and in other variational quantum algorithms. %This is fundamentally about the difficulty of obtaining an appropriate Ansatz that is expressible enough, and this problem is not unique to QAS, in fact being shared with all the other variational algorithms. 
This problem and potential methods to deal with it are further elaborated in Appendix \ref{appendix:expressibility} and \ref{appendix:scaling}. }

Another problem that QAS shares with VQS is that both algorithms require to calculate the pseudo-inverse of a matrix, of whose elements are measured on a quantum computer. This procedure, via singular value decomposition, can be numerically unstable and sensitive to noise, especially with increasing number of ansatz states. In our own simulations, we noticed that for a 5-qubit system, once we simulated shot noise, both algorithms broke down very quickly in general, and this problem would only be further compounded if we ran it on a real quantum computer with all the other sources of noise inherent in them. More studies on the stability problems in finding the pseudo-inverse of the matrix should be conducted to see if this problem will permanently plague VQS and QAS approaches to simulating the dynamics of quantum systems.

\textcolor{Change}{We believe that the QAS method will potentially have the biggest advantage over other quantum and classical algorithms when the following three conditions are fulfilled:} 

\begin{enumerate}
    \item \textcolor{Change}{The basis states which are used to represent the initial quantum state are highly entangled. This will make calculation of corresponding overlaps classically hard, as it boils down to a circuit sampling task. Note that the Quantum Threshold Assumption (QUATH) by Aaronson and Chen \cite{aaronson2016complexity} says that there is no polynomial-time classical algorithm which takes as input a random circuit $C$ and can decide with success probability at least $\frac{1}{2}+\Omega\left(\frac{1}{2^n}\right)$ whether $\vert\langle0^n \vert C \vert0^n \rangle\vert^2$ is greater than the median of $\vert\langle0^n \vert C \vert x^n \rangle\vert^2$ taken over all bit strings $x^n$. In other words, the circuit sampling task is difficult and hence classical algorithms will not be able to compete with algorithms based on circuit sampling as system size scales. The quantum part of QAS is based on circuit sampling which is classically difficult.}
    \item \textcolor{Change}{Hamiltonian should be a linear combination of small number of unitaries or the set of basis states corresponding to the terms in the Hamiltonian should close fast or the Hamiltonian is a low rank matrix. We demonstrate such a case for a Hamiltonian of random Pauli strings in Fig.\ref{fig:multiqubit} for thousands of qubits. By using highly entangled states instead of product states as Ansatz, the evolution is intractable to simulate with classical computers. %As example, for the Hamiltonian of the form $H= XYZXYZXYZ + YZXYZXYZX + ZXYZXYZXY + XXXXXXXX$, the set spanned by the terms in the Hamiltonian is only size 8, implying we can only generate 8 terms in our Ansatz no matter how high our K is. 
    QAS can easily simulate these Hamiltonians, requiring only a limited number of overlaps. However, evolution by Trotterized methods or other variational methods on a quantum computer would be very challenging as it requires deep circuits and many variational parameters. It would even be a challenge to come up with an expressible enough Ansatz for variational methods.
    }
    \item \textcolor{Change}{The system size of interest should be beyond the reach of classical simulation methods.}
\end{enumerate}

\medskip
{\noindent {\em Acknowledgements---}}
% \section{Acknowledgements}
We are grateful to the National Research Foundation and the Ministry of Education, Singapore for financial support. The authors acknowledge the use of the IBM Quantum Experience devices for this work. This work is supported by a Samsung GRC project and the UK Hub in Quantum Computing and Simulation, part of the UK National Quantum Technologies Programme with funding from UKRI EPSRC grant EP/T001062/1.

\bibliographystyle{apsrev4-1}
\bibliography{TDQAS}

\begin{thebibliography}{52}%
\makeatletter
\providecommand \@ifxundefined [1]{%
 \@ifx{#1\undefined}
}%
\providecommand \@ifnum [1]{%
 \ifnum #1\expandafter \@firstoftwo
 \else \expandafter \@secondoftwo
 \fi
}%
\providecommand \@ifx [1]{%
 \ifx #1\expandafter \@firstoftwo
 \else \expandafter \@secondoftwo
 \fi
}%
\providecommand \natexlab [1]{#1}%
\providecommand \enquote  [1]{``#1''}%
\providecommand \bibnamefont  [1]{#1}%
\providecommand \bibfnamefont [1]{#1}%
\providecommand \citenamefont [1]{#1}%
\providecommand \href@noop [0]{\@secondoftwo}%
\providecommand \href [0]{\begingroup \@sanitize@url \@href}%
\providecommand \@href[1]{\@@startlink{#1}\@@href}%
\providecommand \@@href[1]{\endgroup#1\@@endlink}%
\providecommand \@sanitize@url [0]{\catcode `\\12\catcode `\$12\catcode
  `\&12\catcode `\#12\catcode `\^12\catcode `\_12\catcode `\%12\relax}%
\providecommand \@@startlink[1]{}%
\providecommand \@@endlink[0]{}%
\providecommand \url  [0]{\begingroup\@sanitize@url \@url }%
\providecommand \@url [1]{\endgroup\@href {#1}{\urlprefix }}%
\providecommand \urlprefix  [0]{URL }%
\providecommand \Eprint [0]{\href }%
\providecommand \doibase [0]{http://dx.doi.org/}%
\providecommand \selectlanguage [0]{\@gobble}%
\providecommand \bibinfo  [0]{\@secondoftwo}%
\providecommand \bibfield  [0]{\@secondoftwo}%
\providecommand \translation [1]{[#1]}%
\providecommand \BibitemOpen [0]{}%
\providecommand \bibitemStop [0]{}%
\providecommand \bibitemNoStop [0]{.\EOS\space}%
\providecommand \EOS [0]{\spacefactor3000\relax}%
\providecommand \BibitemShut  [1]{\csname bibitem#1\endcsname}%
\let\auto@bib@innerbib\@empty
%</preamble>
\bibitem [{\citenamefont {Preskill}(2018)}]{preskill2018quantum}%
  \BibitemOpen
  \bibfield  {author} {\bibinfo {author} {\bibfnamefont {J.}~\bibnamefont
  {Preskill}},\ }\href@noop {} {\bibfield  {journal} {\bibinfo  {journal}
  {Quantum}\ }\textbf {\bibinfo {volume} {2}},\ \bibinfo {pages} {79} (\bibinfo
  {year} {2018})}\BibitemShut {NoStop}%
\bibitem [{\citenamefont {Deutsch}(2020)}]{deutsch2020harnessing}%
  \BibitemOpen
  \bibfield  {author} {\bibinfo {author} {\bibfnamefont {I.~H.}\ \bibnamefont
  {Deutsch}},\ }\href@noop {} {\bibfield  {journal} {\bibinfo  {journal} {arXiv
  preprint arXiv:2010.10283}\ } (\bibinfo {year} {2020})}\BibitemShut {NoStop}%
\bibitem [{\citenamefont {Peruzzo}\ \emph {et~al.}(2014)\citenamefont
  {Peruzzo}, \citenamefont {McClean}, \citenamefont {Shadbolt}, \citenamefont
  {Yung}, \citenamefont {Zhou}, \citenamefont {Love}, \citenamefont
  {Aspuru-Guzik},\ and\ \citenamefont {Obrien}}]{peruzzo2014variational}%
  \BibitemOpen
  \bibfield  {author} {\bibinfo {author} {\bibfnamefont {A.}~\bibnamefont
  {Peruzzo}}, \bibinfo {author} {\bibfnamefont {J.}~\bibnamefont {McClean}},
  \bibinfo {author} {\bibfnamefont {P.}~\bibnamefont {Shadbolt}}, \bibinfo
  {author} {\bibfnamefont {M.-H.}\ \bibnamefont {Yung}}, \bibinfo {author}
  {\bibfnamefont {X.-Q.}\ \bibnamefont {Zhou}}, \bibinfo {author}
  {\bibfnamefont {P.~J.}\ \bibnamefont {Love}}, \bibinfo {author}
  {\bibfnamefont {A.}~\bibnamefont {Aspuru-Guzik}}, \ and\ \bibinfo {author}
  {\bibfnamefont {J.~L.}\ \bibnamefont {Obrien}},\ }\href@noop {} {\bibfield
  {journal} {\bibinfo  {journal} {Nature communications}\ }\textbf {\bibinfo
  {volume} {5}},\ \bibinfo {pages} {4213} (\bibinfo {year} {2014})}\BibitemShut
  {NoStop}%
\bibitem [{\citenamefont {Quantum}\ \emph {et~al.}(2020)\citenamefont {Quantum}
  \emph {et~al.}}]{google2020hartree}%
  \BibitemOpen
  \bibfield  {author} {\bibinfo {author} {\bibfnamefont {G.~A.}\ \bibnamefont
  {Quantum}} \emph {et~al.},\ }\href@noop {} {\bibfield  {journal} {\bibinfo
  {journal} {Science}\ }\textbf {\bibinfo {volume} {369}},\ \bibinfo {pages}
  {1084} (\bibinfo {year} {2020})}\BibitemShut {NoStop}%
\bibitem [{\citenamefont {Elfving}\ \emph {et~al.}(2020)\citenamefont
  {Elfving}, \citenamefont {Broer}, \citenamefont {Webber}, \citenamefont
  {Gavartin}, \citenamefont {Halls}, \citenamefont {Lorton},\ and\
  \citenamefont {Bochevarov}}]{elfving2020will}%
  \BibitemOpen
  \bibfield  {author} {\bibinfo {author} {\bibfnamefont {V.~E.}\ \bibnamefont
  {Elfving}}, \bibinfo {author} {\bibfnamefont {B.~W.}\ \bibnamefont {Broer}},
  \bibinfo {author} {\bibfnamefont {M.}~\bibnamefont {Webber}}, \bibinfo
  {author} {\bibfnamefont {J.}~\bibnamefont {Gavartin}}, \bibinfo {author}
  {\bibfnamefont {M.~D.}\ \bibnamefont {Halls}}, \bibinfo {author}
  {\bibfnamefont {K.~P.}\ \bibnamefont {Lorton}}, \ and\ \bibinfo {author}
  {\bibfnamefont {A.}~\bibnamefont {Bochevarov}},\ }\href@noop {} {\bibfield
  {journal} {\bibinfo  {journal} {arXiv preprint arXiv:2009.12472}\ } (\bibinfo
  {year} {2020})}\BibitemShut {NoStop}%
\bibitem [{\citenamefont {McClean}\ \emph {et~al.}(2018)\citenamefont
  {McClean}, \citenamefont {Boixo}, \citenamefont {Smelyanskiy}, \citenamefont
  {Babbush},\ and\ \citenamefont {Neven}}]{mcclean2018barren}%
  \BibitemOpen
  \bibfield  {author} {\bibinfo {author} {\bibfnamefont {J.~R.}\ \bibnamefont
  {McClean}}, \bibinfo {author} {\bibfnamefont {S.}~\bibnamefont {Boixo}},
  \bibinfo {author} {\bibfnamefont {V.~N.}\ \bibnamefont {Smelyanskiy}},
  \bibinfo {author} {\bibfnamefont {R.}~\bibnamefont {Babbush}}, \ and\
  \bibinfo {author} {\bibfnamefont {H.}~\bibnamefont {Neven}},\ }\href@noop {}
  {\bibfield  {journal} {\bibinfo  {journal} {Nature communications}\ }\textbf
  {\bibinfo {volume} {9}},\ \bibinfo {pages} {4812} (\bibinfo {year}
  {2018})}\BibitemShut {NoStop}%
\bibitem [{\citenamefont {Huang}\ \emph {et~al.}(2019)\citenamefont {Huang},
  \citenamefont {Bharti},\ and\ \citenamefont {Rebentrost}}]{huang2019near}%
  \BibitemOpen
  \bibfield  {author} {\bibinfo {author} {\bibfnamefont {H.-Y.}\ \bibnamefont
  {Huang}}, \bibinfo {author} {\bibfnamefont {K.}~\bibnamefont {Bharti}}, \
  and\ \bibinfo {author} {\bibfnamefont {P.}~\bibnamefont {Rebentrost}},\
  }\href@noop {} {\bibfield  {journal} {\bibinfo  {journal} {arXiv preprint
  arXiv:1909.07344}\ } (\bibinfo {year} {2019})}\BibitemShut {NoStop}%
\bibitem [{\citenamefont {Cerezo}\ \emph {et~al.}(2020)\citenamefont {Cerezo},
  \citenamefont {Sone}, \citenamefont {Volkoff}, \citenamefont {Cincio},\ and\
  \citenamefont {Coles}}]{cerezo2020cost}%
  \BibitemOpen
  \bibfield  {author} {\bibinfo {author} {\bibfnamefont {M.}~\bibnamefont
  {Cerezo}}, \bibinfo {author} {\bibfnamefont {A.}~\bibnamefont {Sone}},
  \bibinfo {author} {\bibfnamefont {T.}~\bibnamefont {Volkoff}}, \bibinfo
  {author} {\bibfnamefont {L.}~\bibnamefont {Cincio}}, \ and\ \bibinfo {author}
  {\bibfnamefont {P.~J.}\ \bibnamefont {Coles}},\ }\href@noop {} {\bibfield
  {journal} {\bibinfo  {journal} {arXiv preprint arXiv:2001.00550}\ } (\bibinfo
  {year} {2020})}\BibitemShut {NoStop}%
\bibitem [{\citenamefont {Wang}\ \emph {et~al.}(2020)\citenamefont {Wang},
  \citenamefont {Fontana}, \citenamefont {Cerezo}, \citenamefont {Sharma},
  \citenamefont {Sone}, \citenamefont {Cincio},\ and\ \citenamefont
  {Coles}}]{wang2020noise}%
  \BibitemOpen
  \bibfield  {author} {\bibinfo {author} {\bibfnamefont {S.}~\bibnamefont
  {Wang}}, \bibinfo {author} {\bibfnamefont {E.}~\bibnamefont {Fontana}},
  \bibinfo {author} {\bibfnamefont {M.}~\bibnamefont {Cerezo}}, \bibinfo
  {author} {\bibfnamefont {K.}~\bibnamefont {Sharma}}, \bibinfo {author}
  {\bibfnamefont {A.}~\bibnamefont {Sone}}, \bibinfo {author} {\bibfnamefont
  {L.}~\bibnamefont {Cincio}}, \ and\ \bibinfo {author} {\bibfnamefont {P.~J.}\
  \bibnamefont {Coles}},\ }\href@noop {} {\bibfield  {journal} {\bibinfo
  {journal} {arXiv preprint arXiv:2007.14384}\ } (\bibinfo {year}
  {2020})}\BibitemShut {NoStop}%
\bibitem [{\citenamefont {Kandala}\ \emph {et~al.}(2017)\citenamefont
  {Kandala}, \citenamefont {Mezzacapo}, \citenamefont {Temme}, \citenamefont
  {Takita}, \citenamefont {Brink}, \citenamefont {Chow},\ and\ \citenamefont
  {Gambetta}}]{kandala2017hardware}%
  \BibitemOpen
  \bibfield  {author} {\bibinfo {author} {\bibfnamefont {A.}~\bibnamefont
  {Kandala}}, \bibinfo {author} {\bibfnamefont {A.}~\bibnamefont {Mezzacapo}},
  \bibinfo {author} {\bibfnamefont {K.}~\bibnamefont {Temme}}, \bibinfo
  {author} {\bibfnamefont {M.}~\bibnamefont {Takita}}, \bibinfo {author}
  {\bibfnamefont {M.}~\bibnamefont {Brink}}, \bibinfo {author} {\bibfnamefont
  {J.~M.}\ \bibnamefont {Chow}}, \ and\ \bibinfo {author} {\bibfnamefont
  {J.~M.}\ \bibnamefont {Gambetta}},\ }\href@noop {} {\bibfield  {journal}
  {\bibinfo  {journal} {Nature}\ }\textbf {\bibinfo {volume} {549}},\ \bibinfo
  {pages} {242} (\bibinfo {year} {2017})}\BibitemShut {NoStop}%
\bibitem [{\citenamefont {Taube}\ and\ \citenamefont
  {Bartlett}(2006)}]{taube2006new}%
  \BibitemOpen
  \bibfield  {author} {\bibinfo {author} {\bibfnamefont {A.~G.}\ \bibnamefont
  {Taube}}\ and\ \bibinfo {author} {\bibfnamefont {R.~J.}\ \bibnamefont
  {Bartlett}},\ }\href@noop {} {\bibfield  {journal} {\bibinfo  {journal}
  {International journal of quantum chemistry}\ }\textbf {\bibinfo {volume}
  {106}},\ \bibinfo {pages} {3393} (\bibinfo {year} {2006})}\BibitemShut
  {NoStop}%
\bibitem [{\citenamefont {Lee}\ \emph {et~al.}(2018)\citenamefont {Lee},
  \citenamefont {Huggins}, \citenamefont {Head-Gordon},\ and\ \citenamefont
  {Whaley}}]{lee2018generalized}%
  \BibitemOpen
  \bibfield  {author} {\bibinfo {author} {\bibfnamefont {J.}~\bibnamefont
  {Lee}}, \bibinfo {author} {\bibfnamefont {W.~J.}\ \bibnamefont {Huggins}},
  \bibinfo {author} {\bibfnamefont {M.}~\bibnamefont {Head-Gordon}}, \ and\
  \bibinfo {author} {\bibfnamefont {K.~B.}\ \bibnamefont {Whaley}},\
  }\href@noop {} {\bibfield  {journal} {\bibinfo  {journal} {Journal of
  chemical theory and computation}\ }\textbf {\bibinfo {volume} {15}},\
  \bibinfo {pages} {311} (\bibinfo {year} {2018})}\BibitemShut {NoStop}%
\bibitem [{\citenamefont {Bharti}(2020)}]{bharti2020quantum}%
  \BibitemOpen
  \bibfield  {author} {\bibinfo {author} {\bibfnamefont {K.}~\bibnamefont
  {Bharti}},\ }\href@noop {} {\bibfield  {journal} {\bibinfo  {journal} {arXiv
  preprint arXiv:2009.11001}\ } (\bibinfo {year} {2020})}\BibitemShut {NoStop}%
\bibitem [{\citenamefont {Bharti}\ and\ \citenamefont
  {Haug}(2020{\natexlab{a}})}]{bharti2020iterative}%
  \BibitemOpen
  \bibfield  {author} {\bibinfo {author} {\bibfnamefont {K.}~\bibnamefont
  {Bharti}}\ and\ \bibinfo {author} {\bibfnamefont {T.}~\bibnamefont {Haug}},\
  }\href@noop {} {\bibfield  {journal} {\bibinfo  {journal} {arXiv preprint
  arXiv:2010.05638}\ } (\bibinfo {year} {2020}{\natexlab{a}})}\BibitemShut
  {NoStop}%
\bibitem [{\citenamefont {Feynman}(1982)}]{feynman1982simulating}%
  \BibitemOpen
  \bibfield  {author} {\bibinfo {author} {\bibfnamefont {R.~P.}\ \bibnamefont
  {Feynman}},\ }\href@noop {} {\bibfield  {journal} {\bibinfo  {journal} {Int.
  J. Theor. Phys}\ }\textbf {\bibinfo {volume} {21}} (\bibinfo {year}
  {1982})}\BibitemShut {NoStop}%
\bibitem [{\citenamefont {Bharti}\ and\ \citenamefont
  {Haug}(2020{\natexlab{b}})}]{bharti2020quantum2}%
  \BibitemOpen
  \bibfield  {author} {\bibinfo {author} {\bibfnamefont {K.}~\bibnamefont
  {Bharti}}\ and\ \bibinfo {author} {\bibfnamefont {T.}~\bibnamefont {Haug}},\
  }\href@noop {} {\bibfield  {journal} {\bibinfo  {journal} {arXiv preprint
  arXiv:2011.06911}\ } (\bibinfo {year} {2020}{\natexlab{b}})}\BibitemShut
  {NoStop}%
\bibitem [{\citenamefont {Haug}\ and\ \citenamefont
  {Bharti}(2020)}]{haug2020generalized}%
  \BibitemOpen
  \bibfield  {author} {\bibinfo {author} {\bibfnamefont {T.}~\bibnamefont
  {Haug}}\ and\ \bibinfo {author} {\bibfnamefont {K.}~\bibnamefont {Bharti}},\
  }\href@noop {} {\bibfield  {journal} {\bibinfo  {journal} {arXiv preprint
  arXiv:2011.14737}\ } (\bibinfo {year} {2020})}\BibitemShut {NoStop}%
\bibitem [{\citenamefont {Dirac}(1930)}]{dirac1930note}%
  \BibitemOpen
  \bibfield  {author} {\bibinfo {author} {\bibfnamefont {P.~A.}\ \bibnamefont
  {Dirac}},\ }in\ \href@noop {} {\emph {\bibinfo {booktitle} {Mathematical
  Proceedings of the Cambridge Philosophical Society}}},\ Vol.~\bibinfo
  {volume} {26}\ (\bibinfo {organization} {Cambridge University Press},\
  \bibinfo {year} {1930})\ pp.\ \bibinfo {pages} {376--385}\BibitemShut
  {NoStop}%
\bibitem [{\citenamefont {Frenkel}\ \emph {et~al.}(1934)\citenamefont {Frenkel}
  \emph {et~al.}}]{frenkel1934wave}%
  \BibitemOpen
  \bibfield  {author} {\bibinfo {author} {\bibfnamefont {I.}~\bibnamefont
  {Frenkel}} \emph {et~al.},\ }\href@noop {} {\  (\bibinfo {year}
  {1934})}\BibitemShut {NoStop}%
\bibitem [{\citenamefont {McLachlan}(1964)}]{mclachlan1964variational}%
  \BibitemOpen
  \bibfield  {author} {\bibinfo {author} {\bibfnamefont {A.}~\bibnamefont
  {McLachlan}},\ }\href@noop {} {\bibfield  {journal} {\bibinfo  {journal}
  {Molecular Physics}\ }\textbf {\bibinfo {volume} {8}},\ \bibinfo {pages} {39}
  (\bibinfo {year} {1964})}\BibitemShut {NoStop}%
\bibitem [{\citenamefont {Mitarai}\ and\ \citenamefont
  {Fujii}(2019)}]{mitarai2019methodology}%
  \BibitemOpen
  \bibfield  {author} {\bibinfo {author} {\bibfnamefont {K.}~\bibnamefont
  {Mitarai}}\ and\ \bibinfo {author} {\bibfnamefont {K.}~\bibnamefont
  {Fujii}},\ }\href@noop {} {\bibfield  {journal} {\bibinfo  {journal}
  {Physical Review Research}\ }\textbf {\bibinfo {volume} {1}},\ \bibinfo
  {pages} {013006} (\bibinfo {year} {2019})}\BibitemShut {NoStop}%
\bibitem [{\citenamefont {Lloyd}(1996)}]{lloyd1996universal}%
  \BibitemOpen
  \bibfield  {author} {\bibinfo {author} {\bibfnamefont {S.}~\bibnamefont
  {Lloyd}},\ }\href@noop {} {\bibfield  {journal} {\bibinfo  {journal}
  {Science}\ ,\ \bibinfo {pages} {1073}} (\bibinfo {year} {1996})}\BibitemShut
  {NoStop}%
\bibitem [{\citenamefont {Lanyon}\ \emph {et~al.}(2011)\citenamefont {Lanyon},
  \citenamefont {Hempel}, \citenamefont {Nigg}, \citenamefont {M{\"u}ller},
  \citenamefont {Gerritsma}, \citenamefont {Z{\"a}hringer}, \citenamefont
  {Schindler}, \citenamefont {Barreiro}, \citenamefont {Rambach}, \citenamefont
  {Kirchmair} \emph {et~al.}}]{lanyon2011universal}%
  \BibitemOpen
  \bibfield  {author} {\bibinfo {author} {\bibfnamefont {B.~P.}\ \bibnamefont
  {Lanyon}}, \bibinfo {author} {\bibfnamefont {C.}~\bibnamefont {Hempel}},
  \bibinfo {author} {\bibfnamefont {D.}~\bibnamefont {Nigg}}, \bibinfo {author}
  {\bibfnamefont {M.}~\bibnamefont {M{\"u}ller}}, \bibinfo {author}
  {\bibfnamefont {R.}~\bibnamefont {Gerritsma}}, \bibinfo {author}
  {\bibfnamefont {F.}~\bibnamefont {Z{\"a}hringer}}, \bibinfo {author}
  {\bibfnamefont {P.}~\bibnamefont {Schindler}}, \bibinfo {author}
  {\bibfnamefont {J.~T.}\ \bibnamefont {Barreiro}}, \bibinfo {author}
  {\bibfnamefont {M.}~\bibnamefont {Rambach}}, \bibinfo {author} {\bibfnamefont
  {G.}~\bibnamefont {Kirchmair}},  \emph {et~al.},\ }\href@noop {} {\bibfield
  {journal} {\bibinfo  {journal} {Science}\ }\textbf {\bibinfo {volume}
  {334}},\ \bibinfo {pages} {57} (\bibinfo {year} {2011})}\BibitemShut
  {NoStop}%
\bibitem [{\citenamefont {Peng}\ \emph {et~al.}(2005)\citenamefont {Peng},
  \citenamefont {Du},\ and\ \citenamefont {Suter}}]{peng2005quantum}%
  \BibitemOpen
  \bibfield  {author} {\bibinfo {author} {\bibfnamefont {X.}~\bibnamefont
  {Peng}}, \bibinfo {author} {\bibfnamefont {J.}~\bibnamefont {Du}}, \ and\
  \bibinfo {author} {\bibfnamefont {D.}~\bibnamefont {Suter}},\ }\href@noop {}
  {\bibfield  {journal} {\bibinfo  {journal} {Physical review A}\ }\textbf
  {\bibinfo {volume} {71}},\ \bibinfo {pages} {012307} (\bibinfo {year}
  {2005})}\BibitemShut {NoStop}%
\bibitem [{\citenamefont {Barends}\ \emph {et~al.}(2016)\citenamefont
  {Barends}, \citenamefont {Shabani}, \citenamefont {Lamata}, \citenamefont
  {Kelly}, \citenamefont {Mezzacapo}, \citenamefont {Las~Heras}, \citenamefont
  {Babbush}, \citenamefont {Fowler}, \citenamefont {Campbell}, \citenamefont
  {Chen} \emph {et~al.}}]{barends2016digitized}%
  \BibitemOpen
  \bibfield  {author} {\bibinfo {author} {\bibfnamefont {R.}~\bibnamefont
  {Barends}}, \bibinfo {author} {\bibfnamefont {A.}~\bibnamefont {Shabani}},
  \bibinfo {author} {\bibfnamefont {L.}~\bibnamefont {Lamata}}, \bibinfo
  {author} {\bibfnamefont {J.}~\bibnamefont {Kelly}}, \bibinfo {author}
  {\bibfnamefont {A.}~\bibnamefont {Mezzacapo}}, \bibinfo {author}
  {\bibfnamefont {U.}~\bibnamefont {Las~Heras}}, \bibinfo {author}
  {\bibfnamefont {R.}~\bibnamefont {Babbush}}, \bibinfo {author} {\bibfnamefont
  {A.~G.}\ \bibnamefont {Fowler}}, \bibinfo {author} {\bibfnamefont
  {B.}~\bibnamefont {Campbell}}, \bibinfo {author} {\bibfnamefont
  {Y.}~\bibnamefont {Chen}},  \emph {et~al.},\ }\href@noop {} {\bibfield
  {journal} {\bibinfo  {journal} {Nature}\ }\textbf {\bibinfo {volume} {534}},\
  \bibinfo {pages} {222} (\bibinfo {year} {2016})}\BibitemShut {NoStop}%
\bibitem [{\citenamefont {Barends}\ \emph {et~al.}(2015)\citenamefont
  {Barends}, \citenamefont {Lamata}, \citenamefont {Kelly}, \citenamefont
  {Garc{\'\i}a-{\'A}lvarez}, \citenamefont {Fowler}, \citenamefont {Megrant},
  \citenamefont {Jeffrey}, \citenamefont {White}, \citenamefont {Sank},
  \citenamefont {Mutus} \emph {et~al.}}]{barends2015digital}%
  \BibitemOpen
  \bibfield  {author} {\bibinfo {author} {\bibfnamefont {R.}~\bibnamefont
  {Barends}}, \bibinfo {author} {\bibfnamefont {L.}~\bibnamefont {Lamata}},
  \bibinfo {author} {\bibfnamefont {J.}~\bibnamefont {Kelly}}, \bibinfo
  {author} {\bibfnamefont {L.}~\bibnamefont {Garc{\'\i}a-{\'A}lvarez}},
  \bibinfo {author} {\bibfnamefont {A.}~\bibnamefont {Fowler}}, \bibinfo
  {author} {\bibfnamefont {A.}~\bibnamefont {Megrant}}, \bibinfo {author}
  {\bibfnamefont {E.}~\bibnamefont {Jeffrey}}, \bibinfo {author} {\bibfnamefont
  {T.}~\bibnamefont {White}}, \bibinfo {author} {\bibfnamefont
  {D.}~\bibnamefont {Sank}}, \bibinfo {author} {\bibfnamefont {J.}~\bibnamefont
  {Mutus}},  \emph {et~al.},\ }\href@noop {} {\bibfield  {journal} {\bibinfo
  {journal} {Nature communications}\ }\textbf {\bibinfo {volume} {6}},\
  \bibinfo {pages} {1} (\bibinfo {year} {2015})}\BibitemShut {NoStop}%
\bibitem [{\citenamefont {Martinez}\ \emph {et~al.}(2016)\citenamefont
  {Martinez}, \citenamefont {Muschik}, \citenamefont {Schindler}, \citenamefont
  {Nigg}, \citenamefont {Erhard}, \citenamefont {Heyl}, \citenamefont {Hauke},
  \citenamefont {Dalmonte}, \citenamefont {Monz}, \citenamefont {Zoller} \emph
  {et~al.}}]{martinez2016real}%
  \BibitemOpen
  \bibfield  {author} {\bibinfo {author} {\bibfnamefont {E.~A.}\ \bibnamefont
  {Martinez}}, \bibinfo {author} {\bibfnamefont {C.~A.}\ \bibnamefont
  {Muschik}}, \bibinfo {author} {\bibfnamefont {P.}~\bibnamefont {Schindler}},
  \bibinfo {author} {\bibfnamefont {D.}~\bibnamefont {Nigg}}, \bibinfo {author}
  {\bibfnamefont {A.}~\bibnamefont {Erhard}}, \bibinfo {author} {\bibfnamefont
  {M.}~\bibnamefont {Heyl}}, \bibinfo {author} {\bibfnamefont {P.}~\bibnamefont
  {Hauke}}, \bibinfo {author} {\bibfnamefont {M.}~\bibnamefont {Dalmonte}},
  \bibinfo {author} {\bibfnamefont {T.}~\bibnamefont {Monz}}, \bibinfo {author}
  {\bibfnamefont {P.}~\bibnamefont {Zoller}},  \emph {et~al.},\ }\href@noop {}
  {\bibfield  {journal} {\bibinfo  {journal} {Nature}\ }\textbf {\bibinfo
  {volume} {534}},\ \bibinfo {pages} {516} (\bibinfo {year}
  {2016})}\BibitemShut {NoStop}%
\bibitem [{\citenamefont {Sieberer}\ \emph {et~al.}(2019)\citenamefont
  {Sieberer}, \citenamefont {Olsacher}, \citenamefont {Elben}, \citenamefont
  {Heyl}, \citenamefont {Hauke}, \citenamefont {Haake},\ and\ \citenamefont
  {Zoller}}]{sieberer2019digital}%
  \BibitemOpen
  \bibfield  {author} {\bibinfo {author} {\bibfnamefont {L.~M.}\ \bibnamefont
  {Sieberer}}, \bibinfo {author} {\bibfnamefont {T.}~\bibnamefont {Olsacher}},
  \bibinfo {author} {\bibfnamefont {A.}~\bibnamefont {Elben}}, \bibinfo
  {author} {\bibfnamefont {M.}~\bibnamefont {Heyl}}, \bibinfo {author}
  {\bibfnamefont {P.}~\bibnamefont {Hauke}}, \bibinfo {author} {\bibfnamefont
  {F.}~\bibnamefont {Haake}}, \ and\ \bibinfo {author} {\bibfnamefont
  {P.}~\bibnamefont {Zoller}},\ }\href@noop {} {\bibfield  {journal} {\bibinfo
  {journal} {npj Quantum Information}\ }\textbf {\bibinfo {volume} {5}},\
  \bibinfo {pages} {1} (\bibinfo {year} {2019})}\BibitemShut {NoStop}%
\bibitem [{\citenamefont {Poulin}\ \emph {et~al.}(2014)\citenamefont {Poulin},
  \citenamefont {Hastings}, \citenamefont {Wecker}, \citenamefont {Wiebe},
  \citenamefont {Doherty},\ and\ \citenamefont {Troyer}}]{poulin2014trotter}%
  \BibitemOpen
  \bibfield  {author} {\bibinfo {author} {\bibfnamefont {D.}~\bibnamefont
  {Poulin}}, \bibinfo {author} {\bibfnamefont {M.~B.}\ \bibnamefont
  {Hastings}}, \bibinfo {author} {\bibfnamefont {D.}~\bibnamefont {Wecker}},
  \bibinfo {author} {\bibfnamefont {N.}~\bibnamefont {Wiebe}}, \bibinfo
  {author} {\bibfnamefont {A.~C.}\ \bibnamefont {Doherty}}, \ and\ \bibinfo
  {author} {\bibfnamefont {M.}~\bibnamefont {Troyer}},\ }\href@noop {}
  {\bibfield  {journal} {\bibinfo  {journal} {arXiv:1406.4920}\ } (\bibinfo
  {year} {2014})}\BibitemShut {NoStop}%
\bibitem [{\citenamefont {Endo}\ \emph {et~al.}(2020)\citenamefont {Endo},
  \citenamefont {Sun}, \citenamefont {Li}, \citenamefont {Benjamin},\ and\
  \citenamefont {Yuan}}]{endo2020variational}%
  \BibitemOpen
  \bibfield  {author} {\bibinfo {author} {\bibfnamefont {S.}~\bibnamefont
  {Endo}}, \bibinfo {author} {\bibfnamefont {J.}~\bibnamefont {Sun}}, \bibinfo
  {author} {\bibfnamefont {Y.}~\bibnamefont {Li}}, \bibinfo {author}
  {\bibfnamefont {S.~C.}\ \bibnamefont {Benjamin}}, \ and\ \bibinfo {author}
  {\bibfnamefont {X.}~\bibnamefont {Yuan}},\ }\href@noop {} {\bibfield
  {journal} {\bibinfo  {journal} {Phys. Rev. Lett.}\ }\textbf {\bibinfo
  {volume} {125}},\ \bibinfo {pages} {010501} (\bibinfo {year}
  {2020})}\BibitemShut {NoStop}%
\bibitem [{\citenamefont {Yuan}\ \emph {et~al.}(2019)\citenamefont {Yuan},
  \citenamefont {Endo}, \citenamefont {Zhao}, \citenamefont {Li},\ and\
  \citenamefont {Benjamin}}]{yuan2019theory}%
  \BibitemOpen
  \bibfield  {author} {\bibinfo {author} {\bibfnamefont {X.}~\bibnamefont
  {Yuan}}, \bibinfo {author} {\bibfnamefont {S.}~\bibnamefont {Endo}}, \bibinfo
  {author} {\bibfnamefont {Q.}~\bibnamefont {Zhao}}, \bibinfo {author}
  {\bibfnamefont {Y.}~\bibnamefont {Li}}, \ and\ \bibinfo {author}
  {\bibfnamefont {S.~C.}\ \bibnamefont {Benjamin}},\ }\href@noop {} {\bibfield
  {journal} {\bibinfo  {journal} {Quantum}\ }\textbf {\bibinfo {volume} {3}},\
  \bibinfo {pages} {191} (\bibinfo {year} {2019})}\BibitemShut {NoStop}%
\bibitem [{\citenamefont {Li}\ and\ \citenamefont
  {Benjamin}(2017)}]{li2017efficient}%
  \BibitemOpen
  \bibfield  {author} {\bibinfo {author} {\bibfnamefont {Y.}~\bibnamefont
  {Li}}\ and\ \bibinfo {author} {\bibfnamefont {S.~C.}\ \bibnamefont
  {Benjamin}},\ }\href@noop {} {\bibfield  {journal} {\bibinfo  {journal}
  {Physical Review X}\ }\textbf {\bibinfo {volume} {7}},\ \bibinfo {pages}
  {021050} (\bibinfo {year} {2017})}\BibitemShut {NoStop}%
\bibitem [{\citenamefont {Heya}\ \emph {et~al.}(2019)\citenamefont {Heya},
  \citenamefont {Nakanishi}, \citenamefont {Mitarai},\ and\ \citenamefont
  {Fujii}}]{heya2019subspace}%
  \BibitemOpen
  \bibfield  {author} {\bibinfo {author} {\bibfnamefont {K.}~\bibnamefont
  {Heya}}, \bibinfo {author} {\bibfnamefont {K.~M.}\ \bibnamefont {Nakanishi}},
  \bibinfo {author} {\bibfnamefont {K.}~\bibnamefont {Mitarai}}, \ and\
  \bibinfo {author} {\bibfnamefont {K.}~\bibnamefont {Fujii}},\ }\href@noop {}
  {\bibfield  {journal} {\bibinfo  {journal} {arXiv preprint arXiv:1904.08566}\
  } (\bibinfo {year} {2019})}\BibitemShut {NoStop}%
\bibitem [{\citenamefont {Lee}\ \emph {et~al.}(2021)\citenamefont {Lee},
  \citenamefont {Lau}, \citenamefont {Shi},\ and\ \citenamefont
  {Kwek}}]{lee2021simulating}%
  \BibitemOpen
  \bibfield  {author} {\bibinfo {author} {\bibfnamefont {C.-K.}\ \bibnamefont
  {Lee}}, \bibinfo {author} {\bibfnamefont {J.~W.~Z.}\ \bibnamefont {Lau}},
  \bibinfo {author} {\bibfnamefont {L.}~\bibnamefont {Shi}}, \ and\ \bibinfo
  {author} {\bibfnamefont {L.~C.}\ \bibnamefont {Kwek}},\ }\href
  {https://arxiv.com/abs/2101.06879} {\bibfield  {journal} {\bibinfo  {journal}
  {arXiv:2101.06879}\ } (\bibinfo {year} {2021})}\BibitemShut {NoStop}%
\bibitem [{\citenamefont {Yao}\ \emph {et~al.}(2020)\citenamefont {Yao},
  \citenamefont {Gomes}, \citenamefont {Zhang}, \citenamefont {Iadecola},
  \citenamefont {Wang}, \citenamefont {Ho},\ and\ \citenamefont
  {Orth}}]{yao2020adaptive}%
  \BibitemOpen
  \bibfield  {author} {\bibinfo {author} {\bibfnamefont {Y.-X.}\ \bibnamefont
  {Yao}}, \bibinfo {author} {\bibfnamefont {N.}~\bibnamefont {Gomes}}, \bibinfo
  {author} {\bibfnamefont {F.}~\bibnamefont {Zhang}}, \bibinfo {author}
  {\bibfnamefont {T.}~\bibnamefont {Iadecola}}, \bibinfo {author}
  {\bibfnamefont {C.-Z.}\ \bibnamefont {Wang}}, \bibinfo {author}
  {\bibfnamefont {K.-M.}\ \bibnamefont {Ho}}, \ and\ \bibinfo {author}
  {\bibfnamefont {P.~P.}\ \bibnamefont {Orth}},\ }\href@noop {} {\bibfield
  {journal} {\bibinfo  {journal} {arXiv preprint arXiv:2011.00622}\ } (\bibinfo
  {year} {2020})}\BibitemShut {NoStop}%
\bibitem [{\citenamefont {Cirstoiu}\ \emph {et~al.}(2020)\citenamefont
  {Cirstoiu}, \citenamefont {Holmes}, \citenamefont {Iosue}, \citenamefont
  {Cincio}, \citenamefont {Coles},\ and\ \citenamefont
  {Sornborger}}]{cirstoiu2020variational}%
  \BibitemOpen
  \bibfield  {author} {\bibinfo {author} {\bibfnamefont {C.}~\bibnamefont
  {Cirstoiu}}, \bibinfo {author} {\bibfnamefont {Z.}~\bibnamefont {Holmes}},
  \bibinfo {author} {\bibfnamefont {J.}~\bibnamefont {Iosue}}, \bibinfo
  {author} {\bibfnamefont {L.}~\bibnamefont {Cincio}}, \bibinfo {author}
  {\bibfnamefont {P.~J.}\ \bibnamefont {Coles}}, \ and\ \bibinfo {author}
  {\bibfnamefont {A.}~\bibnamefont {Sornborger}},\ }\href@noop {} {\bibfield
  {journal} {\bibinfo  {journal} {npj Quantum Information}\ }\textbf {\bibinfo
  {volume} {6}},\ \bibinfo {pages} {1} (\bibinfo {year} {2020})}\BibitemShut
  {NoStop}%
\bibitem [{\citenamefont {Commeau}\ \emph {et~al.}(2020)\citenamefont
  {Commeau}, \citenamefont {Cerezo}, \citenamefont {Holmes}, \citenamefont
  {Cincio}, \citenamefont {Coles},\ and\ \citenamefont
  {Sornborger}}]{commeau2020variational}%
  \BibitemOpen
  \bibfield  {author} {\bibinfo {author} {\bibfnamefont {B.}~\bibnamefont
  {Commeau}}, \bibinfo {author} {\bibfnamefont {M.}~\bibnamefont {Cerezo}},
  \bibinfo {author} {\bibfnamefont {Z.}~\bibnamefont {Holmes}}, \bibinfo
  {author} {\bibfnamefont {L.}~\bibnamefont {Cincio}}, \bibinfo {author}
  {\bibfnamefont {P.~J.}\ \bibnamefont {Coles}}, \ and\ \bibinfo {author}
  {\bibfnamefont {A.}~\bibnamefont {Sornborger}},\ }\href@noop {} {\bibfield
  {journal} {\bibinfo  {journal} {arXiv preprint arXiv:2009.02559}\ } (\bibinfo
  {year} {2020})}\BibitemShut {NoStop}%
\bibitem [{\citenamefont {Benedetti}\ \emph {et~al.}(2020)\citenamefont
  {Benedetti}, \citenamefont {Fiorentini},\ and\ \citenamefont
  {Lubasch}}]{benedetti2020hardware}%
  \BibitemOpen
  \bibfield  {author} {\bibinfo {author} {\bibfnamefont {M.}~\bibnamefont
  {Benedetti}}, \bibinfo {author} {\bibfnamefont {M.}~\bibnamefont
  {Fiorentini}}, \ and\ \bibinfo {author} {\bibfnamefont {M.}~\bibnamefont
  {Lubasch}},\ }\href@noop {} {\bibfield  {journal} {\bibinfo  {journal} {arXiv
  preprint arXiv:2009.12361}\ } (\bibinfo {year} {2020})}\BibitemShut {NoStop}%
\bibitem [{\citenamefont {d'Alessandro}(2007)}]{d2007introduction}%
  \BibitemOpen
  \bibfield  {author} {\bibinfo {author} {\bibfnamefont {D.}~\bibnamefont
  {d'Alessandro}},\ }\href@noop {} {\emph {\bibinfo {title} {Introduction to
  quantum control and dynamics}}}\ (\bibinfo  {publisher} {CRC press},\
  \bibinfo {year} {2007})\BibitemShut {NoStop}%
\bibitem [{\citenamefont {Dong}\ and\ \citenamefont
  {Petersen}(2010)}]{dong2010quantum}%
  \BibitemOpen
  \bibfield  {author} {\bibinfo {author} {\bibfnamefont {D.}~\bibnamefont
  {Dong}}\ and\ \bibinfo {author} {\bibfnamefont {I.~R.}\ \bibnamefont
  {Petersen}},\ }\href@noop {} {\bibfield  {journal} {\bibinfo  {journal} {IET
  Control Theory \& Applications}\ }\textbf {\bibinfo {volume} {4}},\ \bibinfo
  {pages} {2651} (\bibinfo {year} {2010})}\BibitemShut {NoStop}%
\bibitem [{\citenamefont {Dirr}\ and\ \citenamefont
  {Helmke}(2008)}]{dirr2008lie}%
  \BibitemOpen
  \bibfield  {author} {\bibinfo {author} {\bibfnamefont {G.}~\bibnamefont
  {Dirr}}\ and\ \bibinfo {author} {\bibfnamefont {U.}~\bibnamefont {Helmke}},\
  }\href@noop {} {\bibfield  {journal} {\bibinfo  {journal}
  {GAMM-Mitteilungen}\ }\textbf {\bibinfo {volume} {31}},\ \bibinfo {pages}
  {59} (\bibinfo {year} {2008})}\BibitemShut {NoStop}%
\bibitem [{\citenamefont {Arute}\ \emph {et~al.}(2019)\citenamefont {Arute},
  \citenamefont {Arya}, \citenamefont {Babbush}, \citenamefont {Bacon},
  \citenamefont {Bardin}, \citenamefont {Barends}, \citenamefont {Biswas},
  \citenamefont {Boixo}, \citenamefont {Brandao}, \citenamefont {Buell} \emph
  {et~al.}}]{arute2019quantum}%
  \BibitemOpen
  \bibfield  {author} {\bibinfo {author} {\bibfnamefont {F.}~\bibnamefont
  {Arute}}, \bibinfo {author} {\bibfnamefont {K.}~\bibnamefont {Arya}},
  \bibinfo {author} {\bibfnamefont {R.}~\bibnamefont {Babbush}}, \bibinfo
  {author} {\bibfnamefont {D.}~\bibnamefont {Bacon}}, \bibinfo {author}
  {\bibfnamefont {J.~C.}\ \bibnamefont {Bardin}}, \bibinfo {author}
  {\bibfnamefont {R.}~\bibnamefont {Barends}}, \bibinfo {author} {\bibfnamefont
  {R.}~\bibnamefont {Biswas}}, \bibinfo {author} {\bibfnamefont
  {S.}~\bibnamefont {Boixo}}, \bibinfo {author} {\bibfnamefont {F.~G.}\
  \bibnamefont {Brandao}}, \bibinfo {author} {\bibfnamefont {D.~A.}\
  \bibnamefont {Buell}},  \emph {et~al.},\ }\href@noop {} {\bibfield  {journal}
  {\bibinfo  {journal} {Nature}\ }\textbf {\bibinfo {volume} {574}},\ \bibinfo
  {pages} {505} (\bibinfo {year} {2019})}\BibitemShut {NoStop}%
\bibitem [{\citenamefont {Aaronson}\ and\ \citenamefont
  {Chen}(2016)}]{aaronson2016complexity}%
  \BibitemOpen
  \bibfield  {author} {\bibinfo {author} {\bibfnamefont {S.}~\bibnamefont
  {Aaronson}}\ and\ \bibinfo {author} {\bibfnamefont {L.}~\bibnamefont
  {Chen}},\ }\href@noop {} {\bibfield  {journal} {\bibinfo  {journal} {arXiv
  preprint arXiv:1612.05903}\ } (\bibinfo {year} {2016})}\BibitemShut {NoStop}%
\bibitem [{\citenamefont {Ekert}\ \emph {et~al.}(2002)\citenamefont {Ekert},
  \citenamefont {Alves}, \citenamefont {Oi}, \citenamefont {Horodecki},
  \citenamefont {Horodecki},\ and\ \citenamefont {Kwek}}]{ekert2002direct}%
  \BibitemOpen
  \bibfield  {author} {\bibinfo {author} {\bibfnamefont {A.~K.}\ \bibnamefont
  {Ekert}}, \bibinfo {author} {\bibfnamefont {C.~M.}\ \bibnamefont {Alves}},
  \bibinfo {author} {\bibfnamefont {D.~K.}\ \bibnamefont {Oi}}, \bibinfo
  {author} {\bibfnamefont {M.}~\bibnamefont {Horodecki}}, \bibinfo {author}
  {\bibfnamefont {P.}~\bibnamefont {Horodecki}}, \ and\ \bibinfo {author}
  {\bibfnamefont {L.~C.}\ \bibnamefont {Kwek}},\ }\href@noop {} {\bibfield
  {journal} {\bibinfo  {journal} {Physical review letters}\ }\textbf {\bibinfo
  {volume} {88}},\ \bibinfo {pages} {217901} (\bibinfo {year}
  {2002})}\BibitemShut {NoStop}%
\bibitem [{\citenamefont {Knill}(1995)}]{knill1995approximation}%
  \BibitemOpen
  \bibfield  {author} {\bibinfo {author} {\bibfnamefont {E.}~\bibnamefont
  {Knill}},\ }\href@noop {} {\bibfield  {journal} {\bibinfo  {journal} {arXiv
  preprint quant-ph/9508006}\ } (\bibinfo {year} {1995})}\BibitemShut {NoStop}%
\bibitem [{\citenamefont {M{\"o}tt{\"o}nen}\ \emph {et~al.}(2004)\citenamefont
  {M{\"o}tt{\"o}nen}, \citenamefont {Vartiainen}, \citenamefont {Bergholm},\
  and\ \citenamefont {Salomaa}}]{mottonen2004quantum}%
  \BibitemOpen
  \bibfield  {author} {\bibinfo {author} {\bibfnamefont {M.}~\bibnamefont
  {M{\"o}tt{\"o}nen}}, \bibinfo {author} {\bibfnamefont {J.~J.}\ \bibnamefont
  {Vartiainen}}, \bibinfo {author} {\bibfnamefont {V.}~\bibnamefont
  {Bergholm}}, \ and\ \bibinfo {author} {\bibfnamefont {M.~M.}\ \bibnamefont
  {Salomaa}},\ }\href@noop {} {\bibfield  {journal} {\bibinfo  {journal}
  {Physical review letters}\ }\textbf {\bibinfo {volume} {93}},\ \bibinfo
  {pages} {130502} (\bibinfo {year} {2004})}\BibitemShut {NoStop}%
\bibitem [{\citenamefont {Vartiainen}\ \emph {et~al.}(2004)\citenamefont
  {Vartiainen}, \citenamefont {M{\"o}tt{\"o}nen},\ and\ \citenamefont
  {Salomaa}}]{vartiainen2004efficient}%
  \BibitemOpen
  \bibfield  {author} {\bibinfo {author} {\bibfnamefont {J.~J.}\ \bibnamefont
  {Vartiainen}}, \bibinfo {author} {\bibfnamefont {M.}~\bibnamefont
  {M{\"o}tt{\"o}nen}}, \ and\ \bibinfo {author} {\bibfnamefont {M.~M.}\
  \bibnamefont {Salomaa}},\ }\href@noop {} {\bibfield  {journal} {\bibinfo
  {journal} {Physical review letters}\ }\textbf {\bibinfo {volume} {92}},\
  \bibinfo {pages} {177902} (\bibinfo {year} {2004})}\BibitemShut {NoStop}%
\bibitem [{\citenamefont {Plesch}\ and\ \citenamefont
  {Brukner}(2011)}]{plesch2011quantum}%
  \BibitemOpen
  \bibfield  {author} {\bibinfo {author} {\bibfnamefont {M.}~\bibnamefont
  {Plesch}}\ and\ \bibinfo {author} {\bibfnamefont {{\v{C}}.}~\bibnamefont
  {Brukner}},\ }\href@noop {} {\bibfield  {journal} {\bibinfo  {journal}
  {Physical Review A}\ }\textbf {\bibinfo {volume} {83}},\ \bibinfo {pages}
  {032302} (\bibinfo {year} {2011})}\BibitemShut {NoStop}%
\bibitem [{\citenamefont {Matsuno}(1975)}]{matsuno1975ergodicity}%
  \BibitemOpen
  \bibfield  {author} {\bibinfo {author} {\bibfnamefont {K.}~\bibnamefont
  {Matsuno}},\ }\href@noop {} {\bibfield  {journal} {\bibinfo  {journal}
  {Journal of Mathematical Physics}\ }\textbf {\bibinfo {volume} {16}},\
  \bibinfo {pages} {2368} (\bibinfo {year} {1975})}\BibitemShut {NoStop}%
\bibitem [{\citenamefont {Grant}\ \emph {et~al.}(2019)\citenamefont {Grant},
  \citenamefont {Wossnig}, \citenamefont {Ostaszewski},\ and\ \citenamefont
  {Benedetti}}]{grant2019initialization}%
  \BibitemOpen
  \bibfield  {author} {\bibinfo {author} {\bibfnamefont {E.}~\bibnamefont
  {Grant}}, \bibinfo {author} {\bibfnamefont {L.}~\bibnamefont {Wossnig}},
  \bibinfo {author} {\bibfnamefont {M.}~\bibnamefont {Ostaszewski}}, \ and\
  \bibinfo {author} {\bibfnamefont {M.}~\bibnamefont {Benedetti}},\ }\href@noop
  {} {\bibfield  {journal} {\bibinfo  {journal} {Quantum}\ }\textbf {\bibinfo
  {volume} {3}},\ \bibinfo {pages} {214} (\bibinfo {year} {2019})}\BibitemShut
  {NoStop}%
\bibitem [{\citenamefont {Seki}\ and\ \citenamefont
  {Yunoki}(2021)}]{seki2021quantum}%
  \BibitemOpen
  \bibfield  {author} {\bibinfo {author} {\bibfnamefont {K.}~\bibnamefont
  {Seki}}\ and\ \bibinfo {author} {\bibfnamefont {S.}~\bibnamefont {Yunoki}},\
  }\href@noop {} {\bibfield  {journal} {\bibinfo  {journal} {PRX Quantum}\
  }\textbf {\bibinfo {volume} {2}},\ \bibinfo {pages} {010333} (\bibinfo {year}
  {2021})}\BibitemShut {NoStop}%
\bibitem [{\citenamefont {Abraham}\ \emph {et~al.}(2019)\citenamefont
  {Abraham}, \citenamefont {Akhalwaya}, \citenamefont {Aleksandrowicz},
  \citenamefont {Alexander}, \citenamefont {Alexandrowics}, \citenamefont
  {Arbel}, \citenamefont {Asfaw}, \citenamefont {Azaustre}, \citenamefont
  {Barkoutsos}, \citenamefont {Barron} \emph {et~al.}}]{abraham2019qiskit}%
  \BibitemOpen
  \bibfield  {author} {\bibinfo {author} {\bibfnamefont {H.}~\bibnamefont
  {Abraham}}, \bibinfo {author} {\bibfnamefont {I.~Y.}\ \bibnamefont
  {Akhalwaya}}, \bibinfo {author} {\bibfnamefont {G.}~\bibnamefont
  {Aleksandrowicz}}, \bibinfo {author} {\bibfnamefont {T.}~\bibnamefont
  {Alexander}}, \bibinfo {author} {\bibfnamefont {G.}~\bibnamefont
  {Alexandrowics}}, \bibinfo {author} {\bibfnamefont {E.}~\bibnamefont
  {Arbel}}, \bibinfo {author} {\bibfnamefont {A.}~\bibnamefont {Asfaw}},
  \bibinfo {author} {\bibfnamefont {C.}~\bibnamefont {Azaustre}}, \bibinfo
  {author} {\bibfnamefont {P.}~\bibnamefont {Barkoutsos}}, \bibinfo {author}
  {\bibfnamefont {G.}~\bibnamefont {Barron}},  \emph {et~al.},\ }\href@noop {}
  {\bibfield  {journal} {\bibinfo  {journal} {URL https://qiskit. org}\ }
  (\bibinfo {year} {2019})}\BibitemShut {NoStop}%
\end{thebibliography}%

\newpage
\appendix

\section{Circuits to measure required matrix elements}\label{Appendix_B}
Most applications of Variational quantum algorithms that have been proposed to solve dynamical problems of closed many-body quantum systems require measurements of quantities that are of the form $\text{Re}\bra{\psi}U\ket{\psi}$ and $\text{Im}\bra{\psi}U\ket{\psi}$. This quantity is usually measured with a Hadamard test, which are mostly variations of the procedure proposed in \cite{ekert2002direct}. However, because of the need for a controlled unitary gate (which often needs to act on many qubits), this would be hard for NISQ devices to implement while mantaining high fidelity. If the topology of the quantum computer is linear, this would also necessitate many swap gates to implement if the controlled unitary gate acts on many qubits, which we want to avoid. 

In Variational Quantum Simulation (VQS), where we need to calculate elements like $\frac{\partial \bra{\psi(\Vec{\theta})}}{\partial \theta_i} \frac{\partial \ket{\psi(\Vec{\theta})}}{\partial \theta_j}$ and $\frac{\partial \bra{\psi(\Vec{\theta})}}{\partial \theta_i}H\ket{\psi(\Vec{\theta})}$, the unitary that we need to get the real part of the expectation value usually cannot be expressed as a simple Pauli string. This would then require the Hadamard test mentioned above, which is not good for our fidelity due to the number of gates and the length of the circuit. 

\begin{figure}[H]
    \centering
    \begin{tikzcd}
        \lstick{$\ket{0}$} & \gate{H} & \ctrl{1}&\gate{S^b}&\gate{H}&\meter{}\\
        \lstick{$\ket{\psi}$} &\qwbundle[alternate]{} & \gate{U}\qwbundle[alternate]{}&\qwbundle[alternate]{} &\qwbundle[alternate]{} &\qwbundle[alternate]{}
    \end{tikzcd}
    \caption{Simplest Hadamard test. In the figure, the $S$ gate is the $\sqrt{Z}$ that does a quarter turn around the Bloch sphere. When $b=0$, this measures $\text{Re}\bra{\psi}U\ket{\psi}$, and when $b=1$, this measures $\text{Im}\bra{\psi}U\ket{\psi}$. In general, this controlled unitary can be hard to implement on current noisy devices, especially ones with limited connectivity.}
    \label{hadamardtest}
\end{figure}
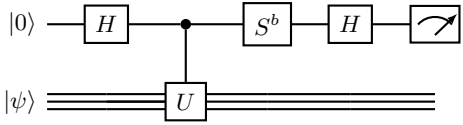

In QAS however, we need to calculate values like $\braket{\chi_i | \chi_j}$ and $\braket{\chi_1 |H| \chi_j}$. They simplify to $\braket{\psi|U_i U_j|\psi}$ and $\braket{\psi|U_i H U_j|\psi}$ respectively. We are usually able to express the unitaries $U_i$, $U_j$ and the Hamiltonian $H$ as Pauli strings. 

Since the product of two Pauli strings would produce another Pauli string up to a factor of $\pm1$ or $\pm i$, these measurement are essentially calculating expectations of a Pauli string. Since this is the case, in QAS we can avoid using the Hadamard test, as this can be easily measured by rotating each qubit into the computational basis corresponding to the Pauli operator.

\section{Detailed description of QAS}\label{Appendix_C}

Here we rederive the results that can be found in \cite{bharti2020quantum2}. We start off with the Schr\"{o}dinger equation:

\begin{gather}
    \frac{\partial \ket{\phi(t)}}{\partial t} = -iH \ket{\phi}.
\end{gather}

If $\ket{\phi(t)}$ is an approximate solution, the statement is no longer true. However, let us define the remainder:

\begin{gather}
    R \equiv \frac{\partial \ket{\phi(t)}}{\partial t} + iH \ket{\phi}.
\end{gather}

	Now we consider the Ansatz:
	\begin{equation}
	    \ket{\phi(\Vec{\alpha}(t))} = \sum_i \alpha_i(t) \ket{\chi_i}.
	\end{equation}

    Normalization of the Ansatz wavefunction is achieved by demanding 
    \begin{gather}
        \Vec{\alpha}^\dagger \mathcal{E}\Vec{\alpha} = 1,\label{normalization}\\
        \mathcal{E}_{i,j} = \braket{\chi_i|\chi_j}.\label{normalization2}
    \end{gather}
    
    The $\ket{\chi_i}$ states are called the K-moment states. 
    
    These states are generated in the following way. If we are considering the Hamiltonian:
    
    \begin{equation}
        H = \sum_j^r \beta_j U_j,
    \end{equation}
    
    where $\beta_j \in \mathbb{C}$ and the $N$ - qubit unitaries $U_j \in SU \left(2^N \equiv \mathcal{N}\right)$, and where each unitary acts non-trivially on at most $\mathcal{O}(poly(logN))$ qubits. If the unitaries are tensored Pauli matrices, we do not need the $\mathcal{O}(poly(logN))$ constraint. We define the $K$-moment states the same way as is defined in \cite{bharti2020quantum2}. Given a set of unitaries (given by the unitary terms of the Hamiltonian) $\mathcal{U}\equiv \{U_i\}_{i=1}^r$ and a positive integer $K$ and some quantum state $\ket{\psi}$, $K$-moment states is the set of quantum states of the form
    
    \begin{equation}
        \{\ket{\chi_i}\}=\{U_{iK}\dots U_{i2}U_{i1}\ket{\psi}\}_i,
    \end{equation}
    
    for $U_{il}\in \mathcal{U}$. This set is denoted by $\mathcal{S}_K$. The cumulative $K$-moment states $\mathcal{CS}_K$ are also defined in \cite{bharti2020quantum2} as $\mathcal{CS}_K \equiv \cup _{j=0}^K \mathcal{S}_j$. Note that if the identity is in $\mathcal{U}$, then $\mathcal{CS}_{K-1} \subseteq \mathcal{S}_K$.
    
    This arbitrary quantum state $\ket{\psi}$ can have variational parameters by itself, i.e. $\ket{\psi}=\ket{\psi(\Vec{\theta})}$, however they are not updated in the QAS procedure and only used for state preparation.
    Note that as extension QAS can subsume variational methods~\cite{bharti2020quantum2}.

Using McLachlan's variational principle \cite{mclachlan1964variational}, we demand the quantity:

\begin{gather}
    I = \int |R|^2 \text{d}v
\end{gather}

on any path must be a minimum for all allowed variations. Now, using the Ansatz proposed in QAS, we obtain:

\begin{gather}
    |R|^2 = \sum_{jk}\Dot{\alpha}^*_j \Dot{\alpha_k} \braket{\chi_j | \chi_k} - i \sum_{jk}\alpha^*_j \Dot{\alpha}_k \braket{\chi_j | H | \chi_k} \notag \\
    + i \sum_{jk} \Dot{\alpha}^*_j \alpha_k \braket{\chi_j | H | \chi_k} + \sum_{jk}\alpha^*_j \alpha_k\braket{\chi_j | H^2 | \chi_k}.
\end{gather}

Now if we vary with respect to $\Dot{\alpha}^*_j$ and $\Dot{\alpha}_j$, and demand that the variations equal to 0, we obtain the condition:

\begin{gather}
    \sum_k \Dot{\alpha}_k \braket{\chi_j|\chi_k} + i \sum_k\alpha_k \braket{\chi_j |H|\chi_k} = 0,
\end{gather}

which is the evolution equation \ref{evolution_eqn}.
%\tfh{Replace bibs with entries from .bib file}
% Now we need a bibliography:

\section{Extension to open systems}\label{opensystemextension}

In a similar manner, the procedure outlined in this paper can be extended to deal with time-dependent Hamiltonians in the open system case. In \cite{haug2020generalized} it is shown that a system interacting with a bath which is described with the Lindblad master equation given as :
    \begin{gather}
        \frac{d}{dt}\rho = -i [H,\rho] + \sum_{n=1}^f \gamma_n (L_n \rho L_n^\dagger - \frac{1}{2}L_n^\dag L_n \rho - \frac{1}{2}\rho L_n^\dag L_n).
    \end{gather}
    Using the ideas in QAS, given a Hilbert space $\mathcal{H}$ and a set of quantum states $\mathcal{S} = \{\ket{\psi_j}\in \mathcal{H}\}_j$, a Hybrid density matrix is introduced as:
    \begin{gather}
    \rho = \sum_{(\ket{\psi_i},\ket{\psi_j})\in\mathcal{S}\cdot \mathcal{S}} \beta_{i,j}\ket{\psi_i}\bra{\psi_j},
    \end{gather}
    where the variational coefficients ($\{\beta_{i,j}\}_{i,j}$) are stored and evolved on a classical device and the quantum states correspond to some quantum system. We can generate the the basis states $\ket{\psi_i}$ with the K-moment expansions, like in QAS. If we introduce the additional notation:
    \begin{gather}
        \mathcal{D}_{i,j} = \bra{\psi_i}H\ket{\psi_j},\\
        \mathcal{R}_{i,j}^n= \bra{\psi_i}L_n\ket{\psi_j},\\
        \mathcal{F}_{i,j}^n= \bra{\psi_i}L_n^\dagger L_n \ket{\psi_j},
    \end{gather}
    the Lindblad master equation can now be written with time dependent parameters $\boldsymbol{\beta}(t)$:
    \begin{gather}
        \mathcal{E} \frac{d}{dt}\boldsymbol{\beta}\mathcal{E} = -i(\mathcal{D}\boldsymbol{\beta}\mathcal{E} - \mathcal{E}\boldsymbol{\beta}\mathcal{D})+ \notag \\
        \sum_{n=1}^f \gamma_n (\mathcal{R}_n \boldsymbol{\beta} R_n^\dagger - \frac{1}{2}\mathcal{F}_n \boldsymbol{\beta} \mathcal{E} - \frac{1}{2}\mathcal{E}\boldsymbol{\beta}\mathcal{F}_n). \label{opensystemeqn}
    \end{gather}
    Once again, the overlaps have to be measured on the quantum computer only once, and all further calculations can be easily performed on a classical computer.

\section{Examples of QAS applied to time-dependent Hamiltonians}\label{QAS_examples}
We now show examples of the QAS algorithm applied to various time-dependent Hamiltonians and initial states. 
    \subsection{1 Qubit Example}
    We first consider the single-qubit Hamiltonian:
    \begin{gather}
        H(t) = Z + \sin(2 \pi t)X.
    \end{gather}
    
    The extended set of permitted operators is then a set of single qubit unitary operations:
    
    \begin{gather}
        S_\text{ext,1} = \{I,X,Y,Z\}.
    \end{gather}
    
\begin{figure}
        \centering
        \includegraphics[width=3in]{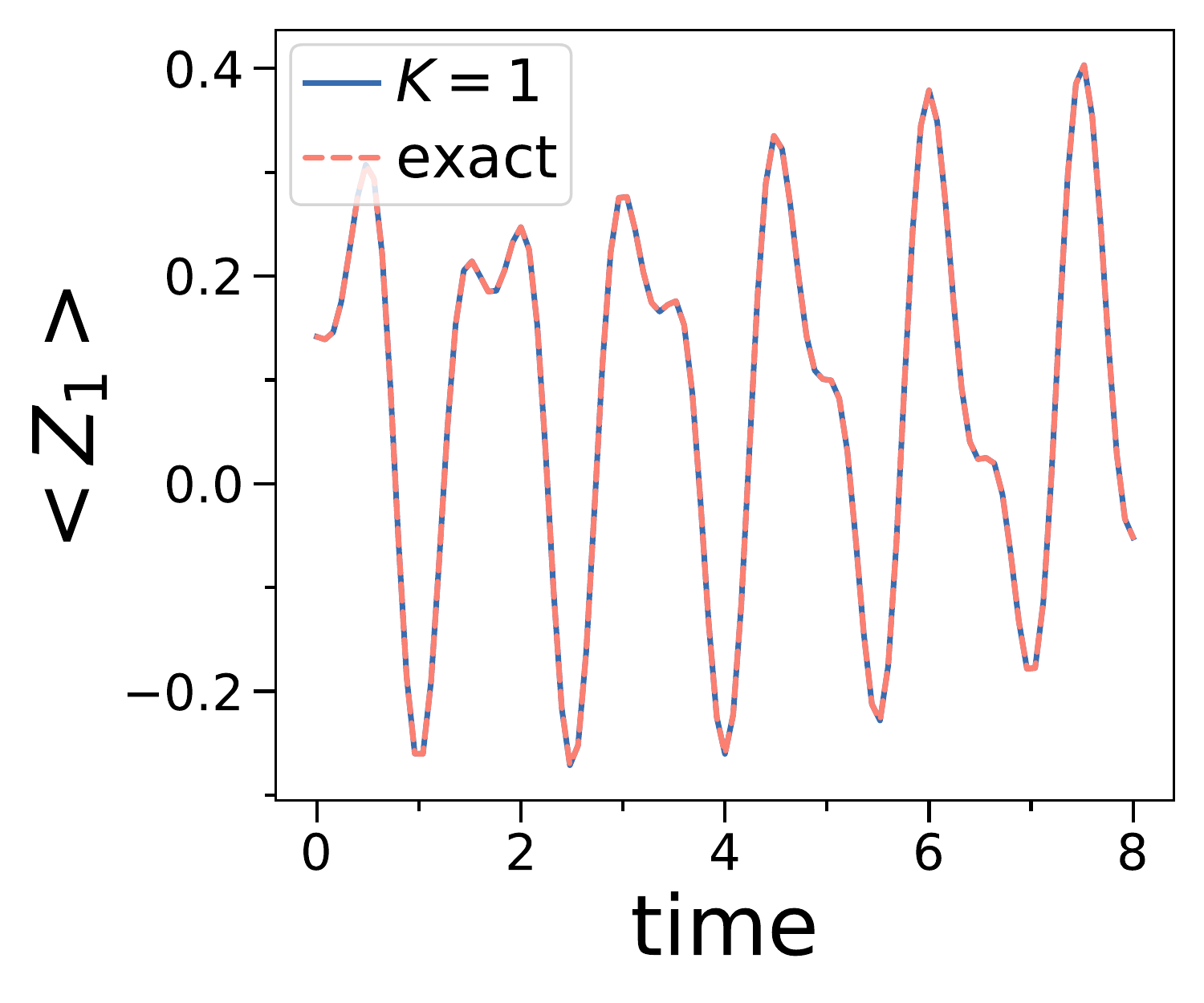}
        \caption{Time evolution of QAS on a single qubit state. Expectation value of $\braket{Z_1}$ with Hamiltonian $ H(t) = Z + \sin(2 \pi t)X $. We show the results for the K=1 moment expansion and the exact result. The fidelity at all times is equal to 1. }
        \label{single_expectation}
    \end{figure}

    With a randomized initial single qubit state state $\ket{\psi_0}$, we construct our Ansatz $\ket{\phi(\Vec{\alpha}(t))}$ as a linear combination of the 1-moment states generated from the $U_i$ in $L_1$:
    
    \begin{gather}
        \ket{\phi(\Vec{\alpha}(t))} \equiv \sum_i \alpha_i(t)\ket{\chi_i}=\sum_i\alpha_i(t)U_i\ket{\psi_0}\notag \\
        =\alpha_0(t)\ket{\psi_0}+ \alpha_1(t)X\ket{\psi_0}
        +\alpha_2(t)Y\ket{\psi_0} + \alpha_3(t)Z\ket{\psi_0}, 
    \end{gather}
    
    while also demanding normalization. In this case, there is no need to investigate higher moment orders as higher orders of K give the exact same set of states to construct our Ansatz with. We then use the quantum computers to calculate the overlap values $\braket{\chi_i|\chi_j}$, $\braket{\chi_i|H_s|\chi_j}$ and $\braket{\chi_i|H_t|\chi_j}$, and use those values to construct our $\mathcal{E}$ and $\mathcal{D}$ matrices. We then evolve the state forward in time by updating the $\alpha$ values. As seen in figure \ref{single_expectation}, the evolution of the state is shown to reproduce the exact behavior very well.
    
    \subsection{3 Qubit Example - No discretization cost}
    We next consider the 3 Qubit example that was discussed above. We start out with the Hamiltonian:
    \begin{gather}
        H(t) = H_s + H_t = (Z_1 Z_2 + Z_2 Z_3) + \sin(2 \pi t) X_2.
    \end{gather}
    This gives us the extended set of permitted operators:
    \begin{gather}
        S_\text{ext,3} = \{I,X_2, Z_1 Z_2, Z_2 Z_3, Z_1 Y_2 , Y_2 Z_3, Z_1 X_2 Z_3\}
    \end{gather}
    
\begin{figure}
        \centering
        \includegraphics[width=3in]{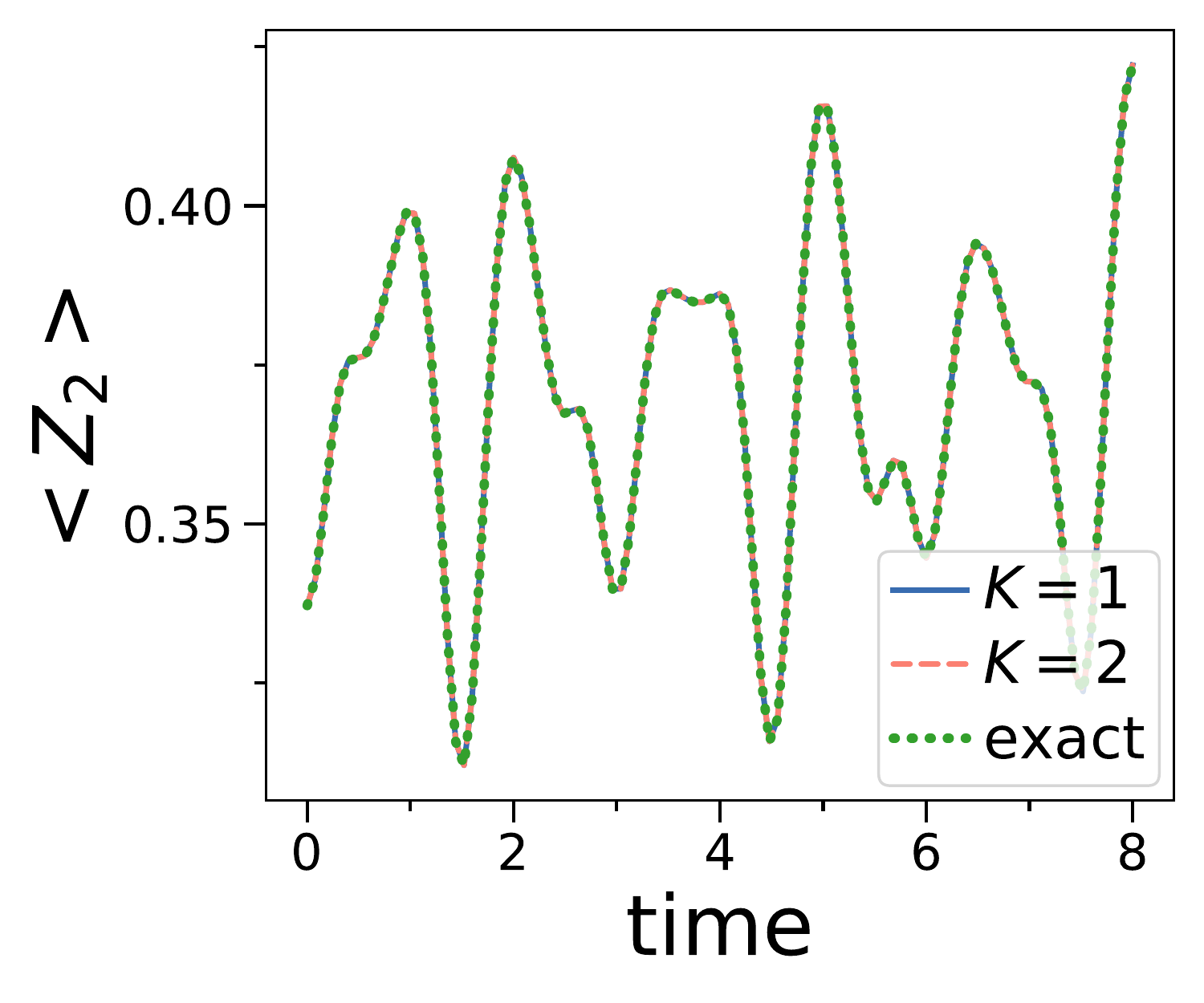}
        \caption{Time evolution of QAS on a 3 qubit state. Expectation value of $\braket{Z_2}$ with Hamiltonian $ H(t) = (Z_1 Z_2 + Z_2 Z_3) + \sin( 2 \pi t) X_2 $. We show the results for the K = 1 and K = 2 moment expansion and the exact result. We use time steps of 0.08 in this example. The fidelity at all times for both moment expansions is equal to 1. }
        \label{three_expectation}
    \end{figure}

    We test this with a randomized three-qubit initial state. In this case, there is only a need to investigate up to 2 moment orders as higher orders of K will give the exact same states as K = 2. As seen in figure \ref{three_expectation}, the evolution of the state reproduces the exact behavior very well again, for both the K = 1 and K = 2 moment states.
    
\begin{figure}
        \centering
        \includegraphics[width=3in]{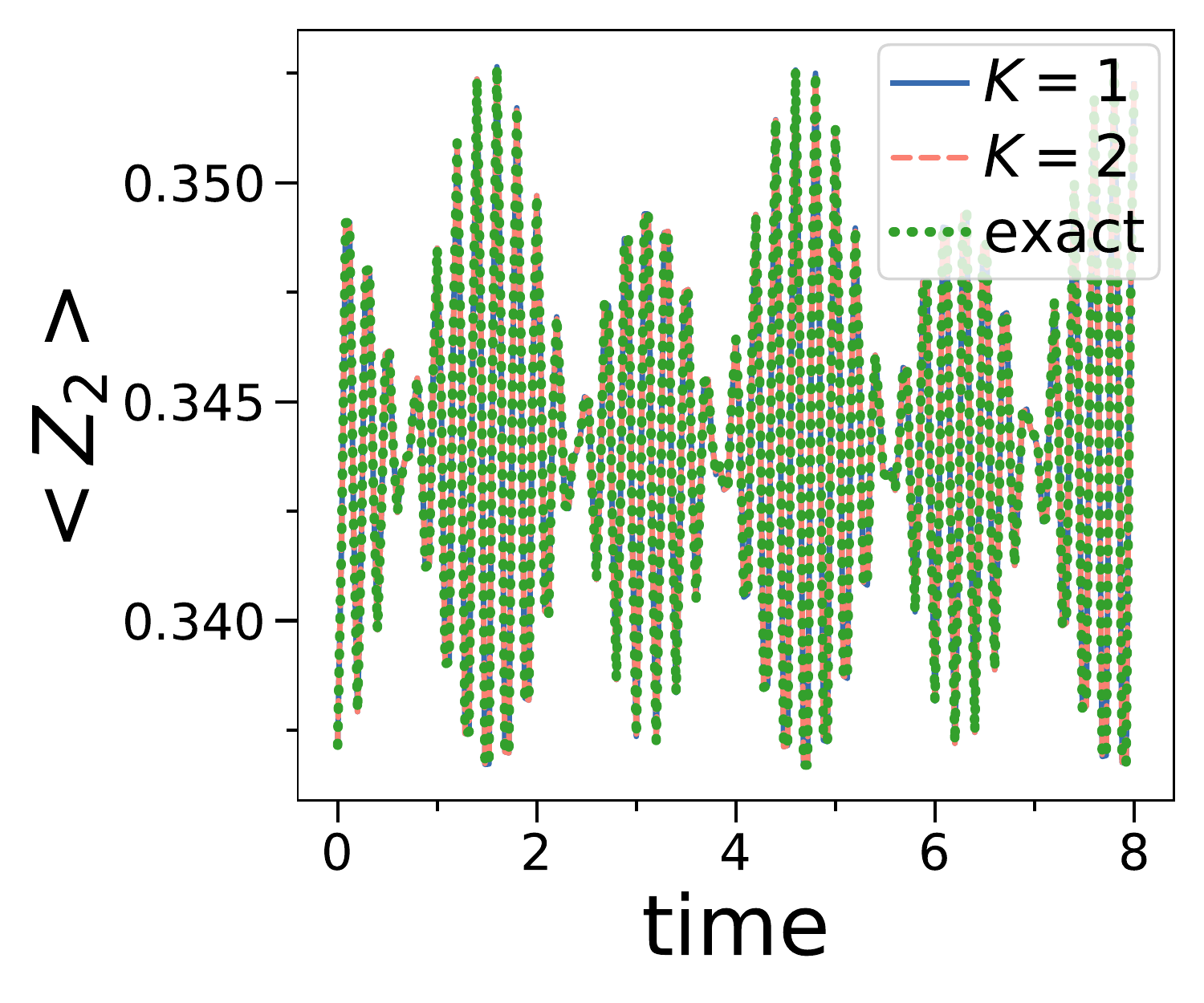}
        \caption{Time evolution of QAS on a 3 qubit state. Expectation value of $\braket{Z_2}$ with Hamiltonian $ H(t) = (Z_1 Z_2 + Z_2 Z_3) + \sin(10 \pi t) X_2 $. We show the results for the K = 1 and K = 2 moment expansion and the exact result. The fidelity at all times for both moment expansions is equal to 1. In this case, we use a time step that is 4 times smaller than the previous case in Figure \ref{three_expectation}, with time step being 0.02. Note that although we use a much finer time step in this example as compared to the previous results, the cost on the quantum computer is still the same as the exact same number of overlap values are calculated on the quantum computer.}
        \label{three_expectation_2}
    \end{figure}

\begin{figure}
        \centering
        \includegraphics[width=3in]{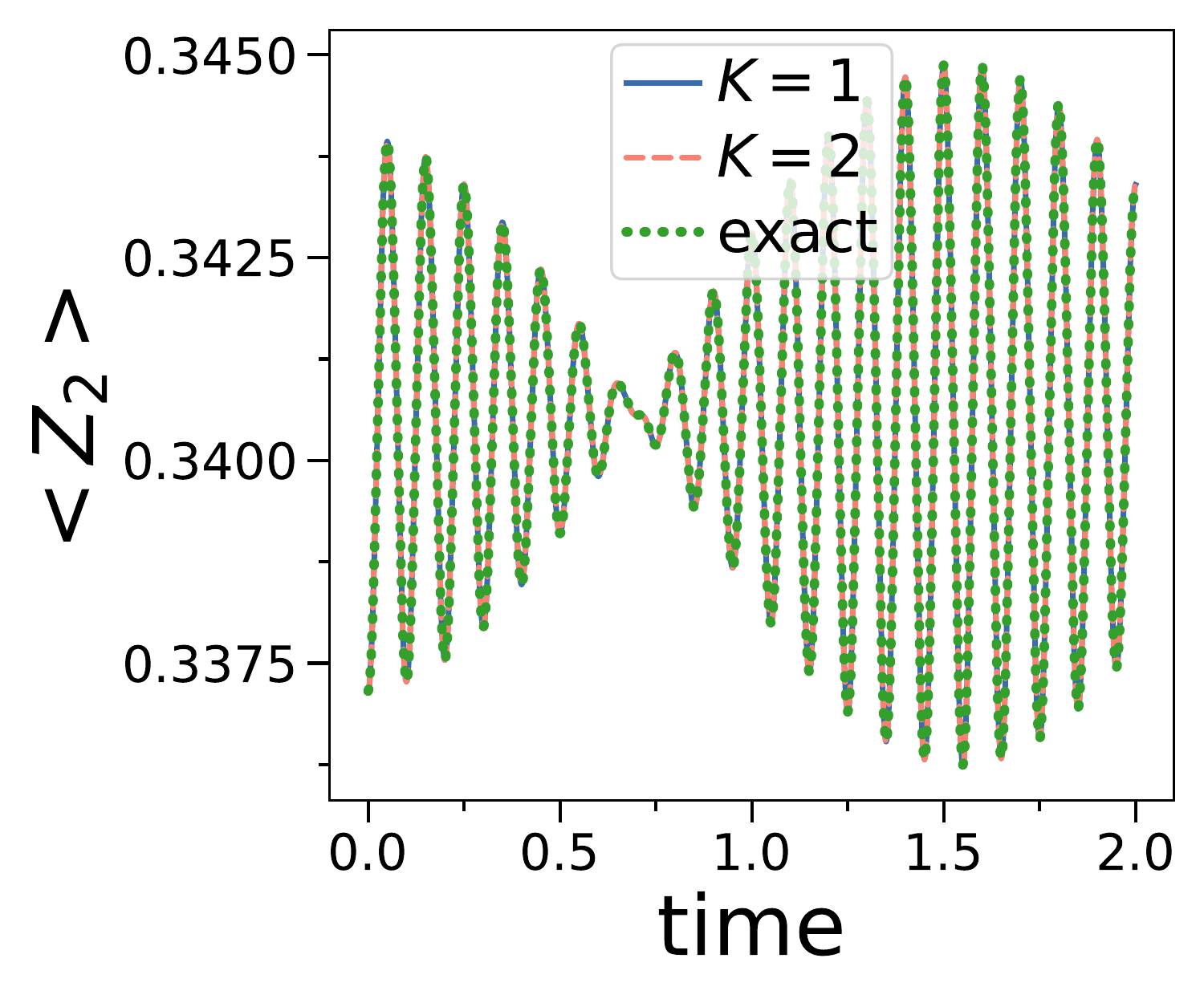}
        \caption{Time evolution of QAS on a 3 qubit state. Expectation value of $\braket{Z_2}$ with Hamiltonian $ H(t) = (Z_1 Z_2 + Z_2 Z_3) + \sin(20 \pi t) X_2 $. We show the results for the K = 1 and K = 2 moment expansion and the exact result. The fidelity at all times for both moment expansions is equal to 1. In this case, we use a time step that is 10 times smaller than the previous case in Figure \ref{three_expectation_2} and 40 times smaller than the case in Figure \ref{three_expectation}, with time step being 0.002. We also simulate it for a shorter time to better see the oscillatory behavior. Again note that although we use a much finer time step in this example as compared to the previous results, the cost on the quantum computer is still the same as the exact same number of overlap values are calculated on the quantum computer.}
        \label{three_expectation_3}
    \end{figure}

    To show that this algorithm does not break down due to discretization problems when we consider higher frequency time-dependent Hamiltonians, as we can arbitrarily decrease the time step as small as we wish with no addition cost on the quantum computer, we also simulate the exact same 3 qubit Hamiltonian, but now instead with frequencies that are higher ($\sin(10 \pi t)$ and $\sin(20 \pi t)$). As seen in Figure \ref{three_expectation_2} and \ref{three_expectation_3}, the algorithm still works for the higher frequencies. In this case, we use a time step that is 4 times and 40 times smaller than the original case to accurately capture the dynamics, but without requiring more circuit evaluations on the quantum computer as exactly the same number of overlap values are calculated on the quantum computer as compared to the earlier case.
    
    \subsection{11 and 7 Qubit Example - Multi-qubit interactions}
    We next want to show that QAS is capable of simulating multiple-qubit interactions easily too. We consider a complex 11 Qubit example. We consider the example of the 1-D Ising model with a time-dependent three body term. The Hamiltonian is given as:
    
    \begin{gather}
        H(t) = \sum_{i=1}^{10}Z_i Z_{i+1} \notag \\
        + \left(\sin(2 \pi t) + \frac{\sin(4 \pi t)}{2}\right)X_5 Y_6 X_7
    \end{gather}
    
\begin{figure}
        \centering
        \includegraphics[width=3in]{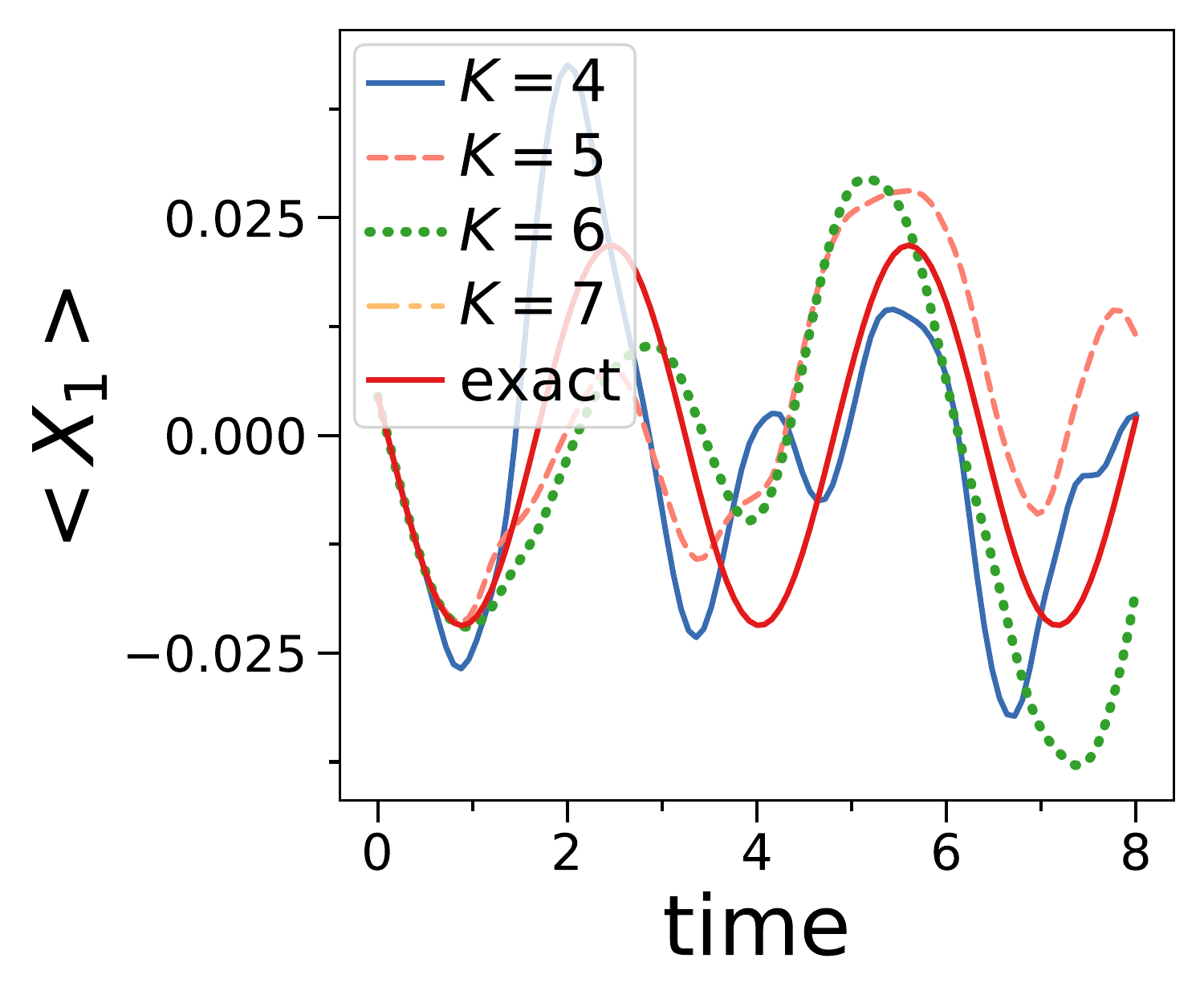}
        \caption{Time evolution of QAS on an 11 qubit state. Expectation value of $\braket{X_1}$. We show the results for the K = 4 to K = 7 moment expansion and the exact result. The K = 7 result is not visible as it lies exactly on the exact result.}
        \label{eleven_expectation}
    \end{figure}

\begin{figure}
        \centering
        \includegraphics[width=3in]{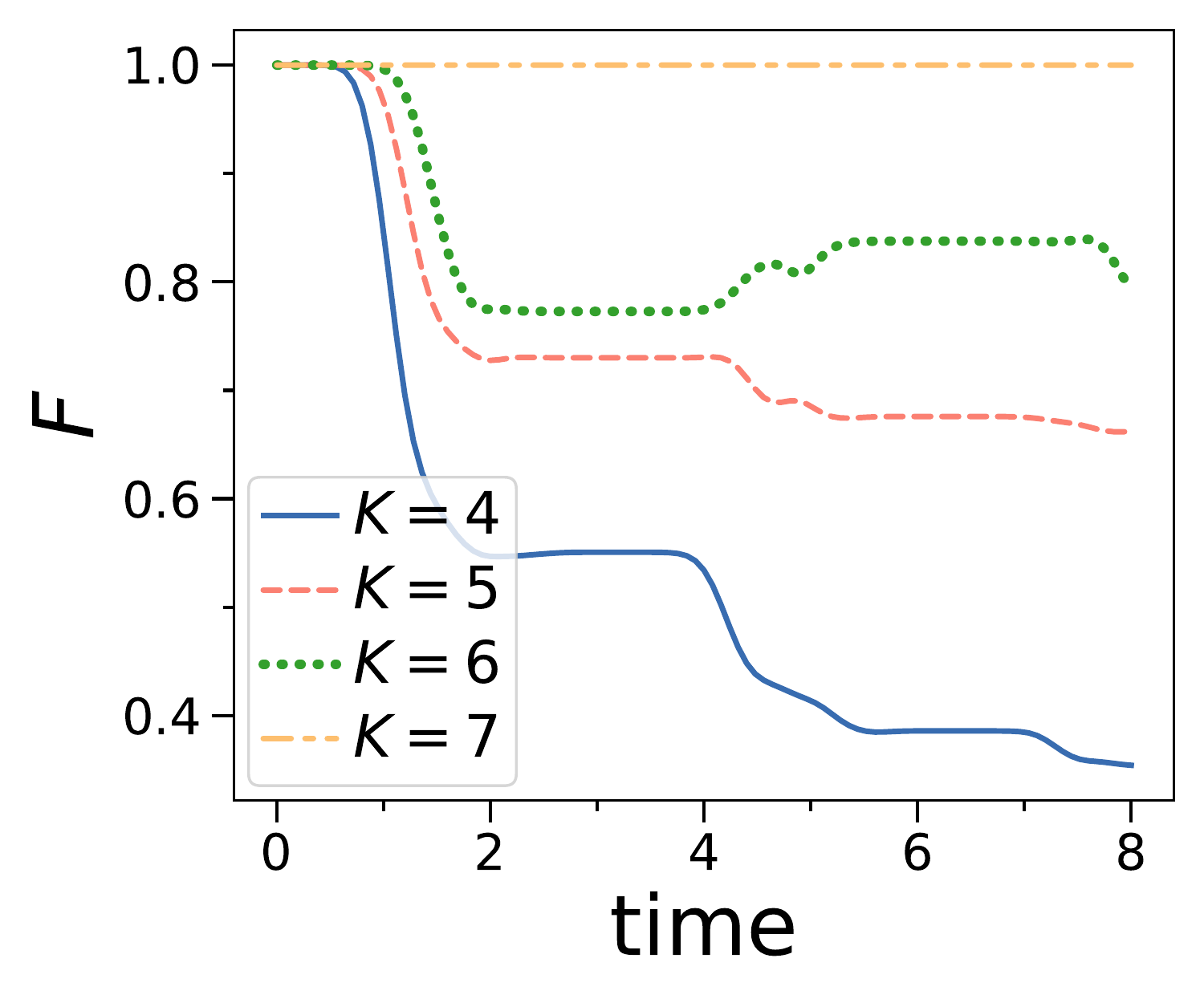}
        \caption{Time evolution of QAS on an 11 qubit state. Fidelity of the state. We show the results for the K = 4 to K = 7 moment expansion and the exact result.}
        \label{eleven_fidelity}
    \end{figure}

    Figure \ref{eleven_expectation} shows the results for the $K$ = 4 to $K$ = 7 expansions ($K$ = 7 expansion is the largest possible expansion, any larger will give the same state space). Figure \ref{eleven_fidelity} shows the fidelity for the expansions. As can be seen, if K is large enough, QAS is able to reproduce the dynamics extremely well. 
    
    We also consider a much more complex 7 Qubit example, with all the terms in the Hamiltonian being three body terms:
    
    \begin{gather}
        H(t) = \sum_{i=1}^{7}Z_iX_{i+1}Z_{i+2} + \sin(2 \pi t)X_3Y_4X_5.
    \end{gather}
    
\begin{figure}
        \centering
        \includegraphics[width=3in]{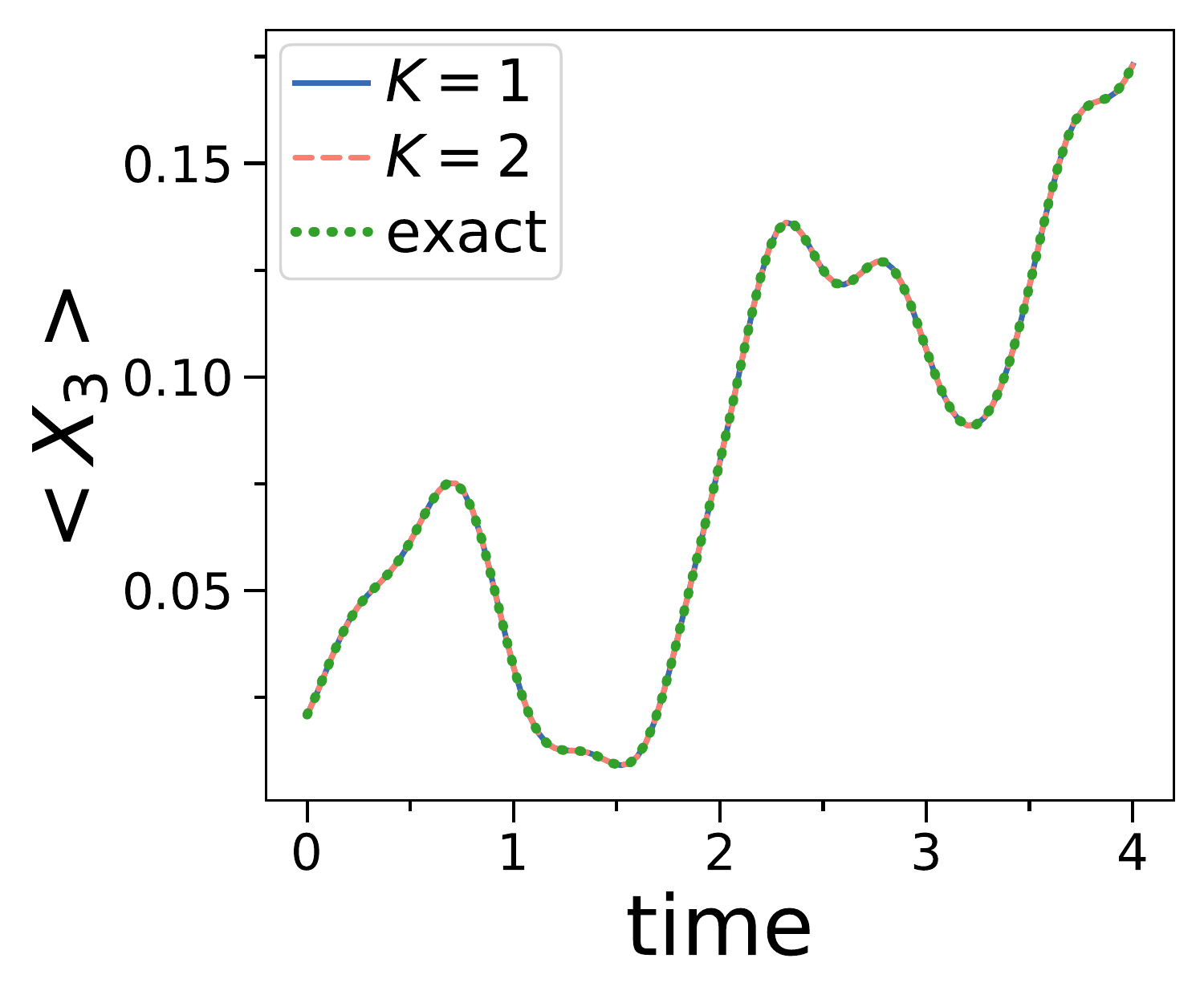}
        \caption{Time evolution of QAS on an 7 qubit state, with Hamiltonian $H(t) = \sum_{i=1}^{7}Z_iX_{i+1}Z_{i+2} + \sin(2 \pi t)X_3Y_4X_5$. Expectation value of $\braket{X_4}$. We show the results for the K = 1 and K = 2 moment expansion and the exact result. The fidelity for both moment expansions is exactly 1.}
        \label{seven_expectation}
    \end{figure}

    The results are show in \ref{seven_expectation}. As can be seen, QAS is able to easily handle evolution under Hamiltonians with many multiple qubit interactions.
    
    % \subsection{6 qubit example - Open systems}
    % Lastly, we consider an example of open system dynamics with a time-dependent Hamiltonian. We consider a six-qubit system under evolution from the Heisenberg Hamiltonian, with an additional $Z_1Z_2$ time dependent term:
    
    % \begin{gather}
    %     H(t) = -\sum_{j=1}^6 \left(X_iX_{i+1} + Y_iY_{i+1} + Z_iZ_{i+1} \right) \notag \\
    %     + \sin (t) Z_1Z_2.
    % \end{gather}
    
    % We use the terms in the Hamiltonian to construct our extended set of permitted operators, and use them to construct our K-moment states in a manner similar to before. For this open system problem, we consider a simple model for decay, with Lindblad terms $L_i = \sqrt{\gamma} \sigma_i^+$. This allows us to construct the matrices necessary for equation \ref{opensystemeqn}. The results are shown in figure \ref{fig:opensystem}. As can be seen, QAS is also able to reproduce the dynamics extremely well even in the open system case, with a time-dependent Hamiltonian.
    
    % \begin{figure}
    %     \centering
    %     \includegraphics[width=3in]{X0_6_time.pdf}
    %     \caption{Time evolution of QAS on a 6 qubit state for an open system. Expectation value of $\braket{X_1}$. We show the result for K=1 and K=2 moment expansions and the exact result. As can be seen, the K=2 result lies exactly on the exact result, with constant fidelity equal to 1. Parameters for the Lindblad term is $\gamma = 1$.}
    %     \label{fig:opensystem}
    % \end{figure}

\section{\textcolor{Change}{Discussion on finding an expressible Ansatz}}\label{appendix:expressibility}

\textcolor{Change}{Given a scalar $\tau$, an $N\times N$ matrix $A$ and
an $N\times1$ vector $v$, the action of the matrix exponential operator
$\exp\left(\tau A\right)$ on $v$ can be approximated as}
\begin{equation}
\exp\left(\tau A\right)v\approx p_{K-1}\left(\tau A\right)v,\label{eq:approx_Krylov}
\end{equation}
\textcolor{Change}{where $p_{K-1}$ is a $K-1$ degree polynomial. The approximation
in equation \ref{eq:approx_Krylov} is an element of the Krylov subspace,}
\begin{equation}
Kr_{K-1}\equiv span\left\{ v,Av,\cdots,A^{K-1}v\right\} .\label{eq:Krylov_1}
\end{equation}
\textcolor{Change}{Thus, the problem of approximating $\exp\left(\tau A\right)v$ can
be recast as finding an element from $Kr_{K}.$ Note that the approximation
in equation \ref{eq:approx_Krylov} becomes exact when $K-1=rank(A).$
In our case, we can identify $v$ with the initial state $\vert\psi\rangle$, $\tau$ with $-\iota t$
and $A$ with the Hamiltonian $H$. }

\textcolor{Change}{In the worst case, the number of overlaps scales as $O(r^K)$ for $r$ terms in $H_{eff}$.
However, depending on the Lie algebra of the Pauli terms in the Hamiltonian and the rank of the Hamiltonian, the number of required overlaps can be a lot smaller compared to the upper bound in the previous sentence.}

\textcolor{Change}{
The number of basis states that was used to construct the hybrid Ansatz, for each $K$ moment expansion, for each Hamiltonian in Appendix \ref{QAS_examples}, is given in Table \ref{tablestates}. }

\begin{table*}[]
\centering
\begin{tabular}{|l|l|l|l|l|l|l|l|}
\hline
              & $K=1$ & $K=2$ & $K=3$ & $K=4$ & $K=5$ & $K=6$ & $K=7$ \\ \hline
1 Qubit Case  & 4     &       &       &       &       &       &       \\ \hline
3 Qubit Case  & 7     & 8     &       &       &       &       &       \\ \hline
7 Qubit Case  & 49    & 64    &       &       &       &       &       \\ \hline
11 Qubit Case & 15    & 92    & 324   & 758   & 1290  & 1724  & 1956  \\ \hline
              &       &       &       &       &       &       &       \\ \hline
\end{tabular}
\caption{\textcolor{Change}{Comparison of the number of basis states used to construct the hybrid Ansatz for each $K$ for each Hamiltonian. For example, the $K=2$ expansion for the 3 qubit case requires 8 quantum states to construct the hybrid Ansatz.}}
\label{tablestates}
\end{table*}

%\textcolor{Change}{At first glance this method is not very promising, as it seems that to get an expressible enough Ansatz to fully capture the dynamics of the quantum state, we would need to consider a $K$ high enough that the number of states in our Ansatz is close to the size of the Hilbert space. This would be a very big limiting factor once we start considering much larger systems. If we have $r$ terms in our $S_{ext}$, we would expect the number of terms to grow as $O(r^K)$ for the first few $K$, as we are essentially building the Krylov subspace.}

\textcolor{Change}{If we have $r$ terms in our $S_{ext}$, the number of terms grows as order $O(r^K)$ for the first few $K$, as we are essentially building the Krylov subspace.}

\textcolor{Change}{This is fundamentally an expressibility problem, present in all NISQ variational algorithms, be it based on linear combination of states or those based on parametric quantum circuits. It is known that to prepare an arbitrary state on an $n$ qubit quantum computer, we require a circuit depth of at least $\frac{1}{n}2^n$~\cite{knill1995approximation,mottonen2004quantum,vartiainen2004efficient,plesch2011quantum}. This suggests that it is very hard to produce an expressible enough Ansatz to reproduce an arbitrary quantum state in the Hilbert space. 
%In most variational algorithms like VQS, we create the Ansatz with parameterized gates. To make it fully expressible, we would need an Ansatz that has an exponential number of parameterized gates. In doing so, the variational algorithms would also be faced with the problem of optimizing over an exponential number of parameters or solving a matrix differential equation, of which the size of the matrix also grows exponentially. In QAS, we construct the Ansatz as a linear combination of quantum states, and similarly, we also expect that to simulate an arbitrary state, we would then need an exponential number of quantum states in our Ansatz.
}

\textcolor{Change}{To reemphasize, all NISQ algorithms (with the exception of Trotter) will definitely face this problem of obtaining an expressible enough Ansatz. If we deal with a Hamiltonian dynamics problem that is ergodic (we expect the quantum state to eventually explore the whole Hilbert space), especially in the long run, we would expect to have to deal with exponential number of parameters. As can be seen in our results, QAS is also not exempt from this limitation in general. And,if we use an arbitrary initial quantum state and an arbitrary Hamiltonian, we would expect it to be ergodic \cite{matsuno1975ergodicity}, which is why in our examples, we see this scaling problem to obtain a good Ansatz.}

\textcolor{Change}{One of the major contributions of the QAS algorithm is that, by using this problem-aware Ansatz, it provides a systematic way to obtain a more and more expressible Ansatz. The other variational algorithms still do not have a systematic method to generate an expressible enough Ansatz, or to improve on an Ansatz in a efficient way. Also, it has been shown that if we use a hardware efficient Ansatz, we would in general expect to encounter the barren plateau problem, which makes it very hard for the algorithm to train and optimize~\cite{mcclean2018barren,grant2019initialization}. Furthermore, the usual technique of using more and more layers of hardware efficient Ansatz circuits gives no guarantee that it will become more and more expressible in an efficient manner, when compared to the number of variational parameters that we are adding. There is also no guarantee that this will indeed improve the appropriateness of the Ansatz. This is especially true for larger systems. In QAS, with the way we generate the Ansatz with $K$ moment states, we can see that at worst, we get an Ansatz with as many states as the size of the Hilbert space, which is fully expressible. This is due to the group of Pauli strings closing on itself eventually. Also, we can see that as we increase the $K$, we will definitely improve our Ansatz and get to a point where it is eventually expressible enough. In future, using the coefficients of the terms in the Hamiltonian, we expect to be able to slow down the growth of the number of states.}

\textcolor{Change}{If we only want to simulate short time evolution, it is sufficient to consider low $K$ and get an expressible enough Ansatz. Figure \ref{eleven_fidelity} shows that for lower $K$, even though for long time periods they are not expressible enough (due to the ergodicity of the Hamiltonian), for short time periods the fidelity is near unity below $t=1$. And we can see that as we increase $K$, the fidelity for longer times systematically increases.}

\textcolor{Change}{In the QAS framework, it might be possible to further reduce the number of states needed to consider by exploiting symmetries in the Hamiltonian, or by choosing a good initial quantum state to generate the K-moment states with, but more work needs to be done in this direction.}

%     \begin{table}[]
%     \centering
% \begin{tabular}{|l|l|l|l|l|}
% \hline
%              & $K=1$ & $K=2$ & $K=3$ & $K=4$ & $K=5$& $K=6$\\ \hline
% 1 Qubit Case & 4     &      &       &       & &\\ \hline
% 3 Qubit Case & 8     & 8     &      &      & &\\ \hline
% 7 Qubit Case & 64     & 64    &       &       & &\\ \hline
% 11 Qubit Case             &       &       &       &       & &\\ \hline
% \end{tabular}
% \caption{Comparison of the number of basis states used to construct the hybrid Ansatz for each $K$ for each Hamiltonian. For example, the $K=2$ expansion for the 4 qubit case, using the Hamiltonian $H_4$, requires 4 quantum states to construct the hybrid Ansatz.}
% \label{tablestates}
% \end{table}

\section{\textcolor{Change}{Scaling analysis}}\label{appendix:scaling}

\textcolor{Change}{As mentioned in the previous Appendix \ref{appendix:expressibility}, our algorithm is closely connected with the idea of generating the Krylov subspace of the Hamiltonian. We use the Krylov subspace to generate our $\ket{\chi}$ states, which are then used to build up our hybrid Ansatz. It is known that the the Krylov subspace spans the entire space when you exponentiate the Hamiltonian $H$ to the power of $K-1$, where $K-1 = rank(H)$. Thus, the number of states that we require in our Ansatz scales linearly with the rank of the Hamiltonian. We believe that this scales favourably compared to other NISQ algorithms previously mentioned in this paper, like VQS and VFF.}

\textcolor{Change}{Our algorithm basically hinges on being able to calculate expectation values of powers of the Hamiltonian, $\braket{\psi|H^k|\psi}$. If we look at the Pauli string level (break our Hamiltonian into linear sums of Pauli strings), the number of Pauli terms in $H^k$ grows exponentially in k. Right now, for current implementation of our algorithm on available quantum computers, this breaking into Pauli strings is necessary due to the imperfections in said quantum computers. However, if we allow more complex operations that cannot be performed very well right now, such as complex controlled unitaries, the resources needed to measure such $\braket{\psi|H^k|\psi}$ values might scale less\cite{seki2021quantum}. }

\section{Circuit for initial state on IBM quantum computer}\label{randomcircuit}

\begin{figure}
    \centering
    \begin{tikzcd}
        \lstick{$\ket{0}$}&\gate{R_x(\Theta_1)}&\gate{R_y(\Theta_2)}&\gate{R_z(\Theta_3)}&\ctrl{1}&\qw& \rstick{$\ket{\phi_0}$}\\
        \lstick{$\ket{0}$}&\gate{R_x(\Theta_4)}&\gate{R_y(\Theta_5)}&\gate{R_z(\Theta_6)}&\gate{Z}&\qw&\rstick{$\ket{\phi_0}$}
    \end{tikzcd}
    \caption{Circuit that generates the initial starting state on the IBM quantum computer for our runs of VQS, QAS and Trotterization. The $\Theta$ were randomly generated and are $[2.846,1.367,3.172,0.011,0.148,0.841]$.}
    \label{fig:random circuit}
\end{figure}
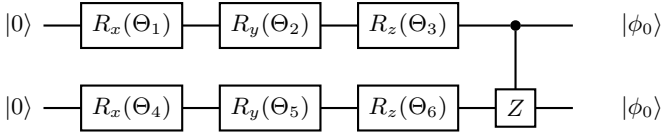

For the runs on the real quantum computer, we generated an initial state with randomized parameters to evolve with the following circuit . It comprised of a $U_3$ rotation with randomized parameters on each qubit, followed by an entangling gate (see Fig.\ref{fig:random circuit}). This initial state is the $\ket{\phi_0}$ that is seen in the circuits \ref{fig:VQS_circuit} and \ref{fig:QAS_circuit}.

\begin{figure}%[htbp]
         \centering
         \includegraphics[width=3.2in]{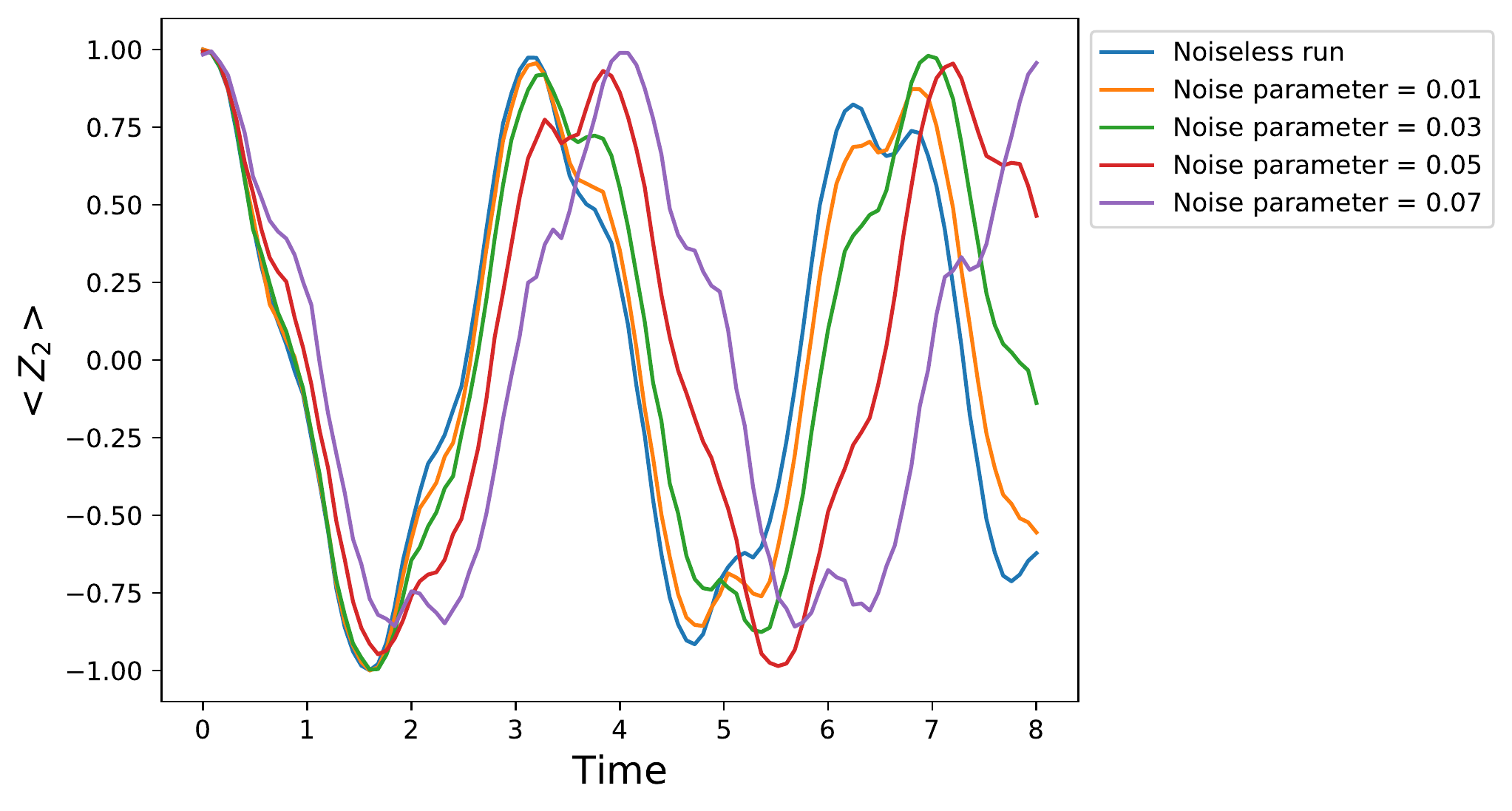}
         \caption{VQS run with depolaring noise for different noise parameters $\lambda$. As can be seen, as the noise parameter increases (simulating more noise in the system), the systematic error in VQS, which is causing the lag in the results for VQS, increases. VQS appears to be sensitive to this systematic error that overestimates periods in the presence of noise.}
         \label{fig:noise}
\end{figure}

\section{\textcolor{Change}{Error mitigation techniques for IBM quantum computer}}\label{appendix:errormitigation}

\textcolor{Change}{We performed the following error mitigation technique on the IBM quantum computer data to mitigate measurement noise using a common method on Qiskit~\cite{abraham2019qiskit}. We calibrated the measurement noise with a calibration matrix $W$. To obtain the matrix $W$ for the 2-qubit case, we first prepare a circuit that starts out with the starting state $\ket{00}$ and measure the probability to obtain each outcome in the measurement basis. We repeat this process for the other 3 possible starting states $\ket{01}$, $\ket{10}$ and $\ket{11}$. By doing so, we can obtain the matrix $W$, defined as $C_{output} = W C_{input}$, where $C$ is the vector representing the state. This $W$ matrix in essence is some form of characterization of the noise from measurement errors. }%We first express the probabilities of obtaining the basis states $\ket{00}$, $\ket{01}$, $\ket{10}$ and $\ket{11}$ from a quantum circuit in a vector $C_{noisy}$.
\textcolor{Change}{Every time we run a circuit and obtain the result $C_{noisy}$, we apply the inverse of the matrix $W$. This relates the probability vector $C_{noisy}$ with the ideal measurement outcome without errors $C_{ideal}$, i.e.  $C_{noisy} = W C_{ideal}$.} %To obtain the matrix $W$ for the 2-qubit case, we independently prepare each of the input states $\ket{00}$, $\ket{01}$, $\ket{10}$ and $\ket{11}$, and measure that probability for each outcome. 
\textcolor{Change}{This was repeated on a daily basis for every day we performed the experiment on the IBM quantum computer. For details, refer to the Qiskit online textbook at~\cite{abraham2019qiskit}.}

\section{Observing the effect of depolarizing noise on VQS}\label{appendix:noise}

We observe how the effects of depolarizing noise contributes to a systematic error in VQS in Fig.\ref{fig:noise}. As can be seen, when we increase the noise parameter $\lambda$ that controls the depolarizing noise, the systematic error or lag in VQS increases. We model the noise as following
\begin{gather}
    E_{\text{depolarizing}}(\rho) = (1-\lambda)\rho + \lambda \text{Tr}[\rho^2]I.
\end{gather}
The depolarizing error can be interpreted as the probability that a single qubit will be replaced by the completely mixed state after being operated on.

\end{document}